\documentclass[amssymb,usenatbib,useAMS,usedcolumns]{mn2e}
\usepackage{times} 
\usepackage{latexsym,amssymb,amsmath,amsfonts}
\usepackage{rotfloat} 
\usepackage{rotating} 
\usepackage{float} 
\usepackage{graphicx}
\usepackage{natbib}
\usepackage{supertabular}
\usepackage{longtable}
\usepackage{url}
\usepackage{dcolumn}
\usepackage{placeins}
\usepackage{multirow} 
\usepackage{rotating}
\begin{document}
\newcommand{\hon}{\mbox{H\,{\sc i}}}
\newcommand{\hto}{\mbox{H\,{\sc ii}}}
\newcommand{\heo}{\mbox{He\,{\sc i}}}
\newcommand{\het}{\mbox{He\,{\sc ii}}}
\newcommand{\oon}{\mbox{O\,{\sc i}}}
\newcommand{\cto}{\mbox{C\,{\sc ii}}}
\newcommand{\con}{\mbox{C\,{\sc i}}}
\newcommand{\sit}{\mbox{Si\,{\sc ii}}}
\newcommand{\sto}{\mbox{S\,{\sc ii}}}
\newcommand{\non}{\mbox{N\,{\sc i}}}
\newcommand{\mgt}{\mbox{Mg\,{\sc ii}}}
\newcommand{\fet}{\mbox{Fe\,{\sc ii}}}
\newcommand{\alt}{\mbox{Al\,{\sc ii}}}
\newcommand{\nit}{\mbox{Ni\,{\sc ii}}}
\newcommand{\cfo}{\mbox{C\,{\sc iv}}}
\newcommand{\sif}{\mbox{Si\,{\sc iv}}}
\newcommand{\sith}{\mbox{Si\,{\sc iii}}}
\newcommand{\alth}{\mbox{Al\,{\sc iii}}}
%
%
\title[Low-metallicity DLAs at high z]{
{A study of low-metallicity DLAs at high redshift and $\cto$* as a probe of their physical conditions
\thanks{Based on observations 383.A-0272 and 086.A-0204 (PI: P. Petitjean) made with the VLT/UVES, which is 
operated by the European Southern Observatory, and archival Keck/HIRES data Prgm. ID. A185Hb (PI: Pettini) and
Prgm. ID. U39H (PI: A. Wolfe) }}  
\author[R. Dutta et al.]{R. Dutta$^{1}$\thanks{E-mail: rdutta@iucaa.ernet.in},  
R. Srianand$^{1}$, H. Rahmani$^{1,2}$, P. Petitjean $^{3}$, P. Noterdaeme $^{3}$, C. Ledoux $^{4}$  \\    
$^{1}$ Inter-University Centre for Astronomy and Astrophysics, Post Bag 4, 
Ganeshkhind, Pune 411\,007, India \\   
$^{2}$ School of Astronomy, Institute for Research in Fundamental Sciences (IPM), PO Box 19395-5531, Tehran, Iran \\
$^{3}$ Institut d'Astrophysique de Paris, CNRS and UPMC Paris 6, UMR7095, 98bis boulevard Arago, 75014 Paris, France \\
$^{4}$ European Southern Observatory, Alonso de C\'{o}rdova 3107, Casilla 19001, Vitacura, Santiago, Chile \\
} 
}
\date{Accepted. Received; in original form }
\maketitle
\label{firstpage}
%
%
\begin {abstract}  
\noindent We present a detailed high spectral resolution (R $\sim$ 40000) study of five high-z damped
Lyman $\alpha$ systems (DLAs) and one sub-DLA detected along four QSO sightlines. Four of these DLAs are very 
metal-poor with [Fe/H] $\le$ $-$2. One of them, at z$_{abs}$ = 4.20287 towards J0953$-$0504, is the most metal-
poor DLA at z $>$ 4 known till date. This system shows no enhancement of C over Fe and O, and standard Population 
II star yields can explain its relative abundance pattern. The DLA at z$_{abs}$ = 2.34006 towards J0035$-$0918 
has been claimed to be the most carbon-enhanced metal-poor DLA. However, we show that thermal
broadening is dominant in this system and, when this effect is taken into account, the measured 
carbon enhancement ([C/Fe] = 0.45 $\pm$ 0.19) becomes $\sim$ 10 times less than what was reported previously.
The gas temperature in this DLA is estimated to be in the range of 5000 $-$ 8000 K, consistent with a warm 
neutral medium phase. From photoionization modelling of two of the DLAs showing $\cto$* absorption, 
we find that the metagalactic background radiation alone is not sufficient to explain the observed $\cto$* 
cooling rate, and local heating sources, probably produced by in-situ star formation, are needed. Cosmic ray 
heating is found to contribute $\gtrsim$ 60\% to the total heating in these systems. Using a sample
of metal-poor DLAs with $\cto$* measurements, we conclude that the cosmic ray ionization rate is
equal to or greater than that seen in the Milky Way in $\sim$ 33\% of the systems with $\cto$* detections.
\end {abstract}  
%
%
%
%
\section{Introduction} 
\label{sec_intro}  
Understanding the formation and evolution of the first generation of stars would help in fathoming the initial chemical conditions 
and evolution of the universe \citep{brom04}. In addition, it is believed that the radiation from such Population III stars may have 
played an important role in the reionization of the universe. These massive stars emit 10$^{47}$ $-$ 10$^{48}$ $\hon$ and $\heo$ 
ionizing photons per second per solar mass of formed stars, where the lower value holds for stars of $\sim$ 20 M$\odot$ and the upper 
value applies to stars with $\gtrsim$ 100 M$\odot$ \citep[see][]{tum00,brom01}. However, their lifetimes are short ($\sim$ 3 
$\times$ 10$^{6}$ years) and hence it would be very difficult to observe them directly. Nevertheless, we can try to infer
signatures of Population III stars in the most metal-poor environments detected in nature. Recently there have also been 
indications of Population III star formation in the universe, even as late as z $\sim$ 3, possibly due to inefficient transport 
of heavy elements and/or poor mixing that leaves pockets of pristine gas \citep{jim06,tor07,ino11,cas13}. \\
\indent The metal-poor stars in the halo of the Milky Way are being studied in an effort to understand the chemical composition 
and hence the nature of star formation at very early epochs \citep[see][]{akr04,cay04,beer05,freb10}. Such studies can also be 
used to infer the chemical evolution in the universe, if most galaxies follow the chemical history of the Milky Way. However, the 
measurement of abundances in stellar atmospheres is hampered by having to take into account non-local thermodynamic equilibrium 
effects as well as three dimensional effects in one dimensional stellar atmosphere models \citep{asp05}. In addition, the presence of
convection or accretion of material from a companion can affect the photospheric abundance measurements. Hence, these measurements 
may not always trace well the chemical evolution history of the Galaxy. \\
\indent These issues are not present when one studies damped Ly$\alpha$ systems (DLAs), which are by definition, clouds of neutral gas 
with neutral hydrogen column density \textit{N}($\hon$) \textgreater~2 $\times$ 10$^{20}$ $\hon$ atoms cm$^{-2}$ \citep[see for a review]
[]{wolf05}. The high \textit{N}($\hon$) self-shields the gas from the ultraviolet background radiation of quasars (QSOs) and galaxies. 
Due to this, abundance measurements require negligible ionization corrections, i.e., hydrogen gas in DLAs is mostly neutral and most 
of the heavy elements are either neutral or singly ionized \citep{petn02,pro02,cen03,led03}. Even then, line saturation and presence 
of dust may hinder abundance measurements in DLAs. However, when the metallicity of the DLAs happens to be $\lesssim$ 10$^{-2}$ 
Z$_{\odot}$ (i.e. in so called metal-poor DLAs), these issues are also mitigated \citep[for e.g.][]{akr05}, especially for the 
estimation of C, N and O abundances. Moreover, high-z metal-poor DLAs are believed to probe gas in or around protogalaxies, and hence 
contain evidences of the earliest star formation and retain signatures of the initial chemical enrichment. The relative abundances of C, 
N, O, Fe, etc., in low-metallicity DLAs can give insights into the type of stars (specifically the stellar initial mass function, IMF) that 
led to their production, and hence shed light on our understanding of galactic nucleosynthesis and chemical evolution \citep{petj08,petn08, 
penp10,cookb11}. In view of this, we undertake a study of five DLAs (pre-selected as candidate low-metallicity DLAs), and their elemental abundances, 
and compare our results with the existing measurements of low-metallicity DLAs and stars in the Galactic halo.\\
\indent There is an ongoing interest in detecting and studying the properties of metal-poor DLAs. \citet{petn08} studied a sample of four 
metal-poor DLAs (defined as those with [Fe/H] $\le$ $-$2.0) using high resolution spectra (R $\sim$ 40000) from Keck and VLT. They found 
that the C/O ratio trend at low metallicity in high redshift DLAs matches that of halo stars in the Galaxy \citep{akr04,spi05}, and the 
ratio shows higher values that cannot be explained by nucleosynthesis models based on Population II stars. Their results also suggested 
that the N/O ratio at low metallicities may show a minimum value. \citet{penp10} used medium resolution spectra (R $\sim$ 5000) to study 
a sample of 35 DLAs pre-selected as metal-poor from the SDSS DR5 database. However, while medium resolution spectra is sufficient to detect 
metal-poor DLAs, one needs to be cautious about employing them to measure accurate abundances in the most metal-poor systems, which typically 
have line widths $\lesssim$ 10 km s$^{-1}$. \\
\indent Measurements of seven new metal-poor DLAs observed with high resolution spectrograph were reported by \citet{cookb11}, who have also 
compiled a sample of all the (22) metal-poor DLAs known in literature. Using this sample and taking oxygen as the reference element, they have 
defined the typical abundance pattern of a very metal poor (VMP) DLA by determining the mean $\langle$X/O$\rangle$ ratio for each available 
element X, and then referring this mean value to the adopted solar scale. On comparing this with the yields of both Population II and III stars, 
it is found that, while a standard model of Population III stars (a top-heavy IMF where the stars explode as core collapse supernovae) gives 
reasonable agreement with the observed abundance pattern of a typical VMP DLA, the possibility that Population II stars may account for the observed 
abundances cannot be discarded. To distinguish between these two nucleosynthesis models, the [C/Fe] ratio is useful, as current models of Population 
II stars cannot explain an enhanced [C/Fe] ratio (along with [N/Fe]$\lesssim$ 0). An enhanced [O/Fe] ratio ($\gtrsim$ 0.4) will also be an useful 
discriminatory tool, since enhanced [C/Fe] is likely to lead to enhanced [O/Fe]. There are only two examples of carbon enhancement in metal-poor DLAs 
\citep{cooka11,cook12}, which are likely to be the outcome of nucleosynthesis in massive stars. Understanding the abundance pattern in these 
carbon-enhanced metal-poor (CEMP) DLAs and increasing their number is important for understanding the true metal yields from very massive stars. \\
\indent In addition, since DLAs are the main reservoirs of neutral gas at high z \citep{not09a,pro05}, studying their star formation 
properties is important, as they could contribute significantly to the global star formation rate density at high z \citep[for e.g.,][]
{wolf08,rah10}. Detection of H$_{2}$ or $\cto$* absorption in DLAs can allow us to study the physical conditions in the absorbing 
gas, and also in principle to infer the star formation rate \citep[see][]{wolf03a,sri05}. In the present sample, we have observed 
$\cto$* absorption associated with two of the DLAs. The $\cto$* $\lambda$1335 column density has been used to measure the [$\cto$] 
$\lambda$158 $\mu$m cooling rate in the neutral gas. We simulate the physical conditions in these two DLAs, and try to match the 
model predictions with the observations in order to infer the ambient radiation field and cosmic ray ionization rate in these DLAs. 
Further, we look at all the detections of $\cto$* in metal-poor DLAs reported till date and try to relate the $\cto$* absorption to 
the metallicity and \textit{N}($\hon$), in order to understand the physical conditions in such DLAs. \\
\indent This paper is organized as follows. In Section 2, we give details of the observations and data reduction process. How the 
analysis of the data was carried out is explained is Section 3. Section 4 gives description of the individual absorption systems. 
In Section 5, we discuss the trends observed in abundance ratios of elements in metal-poor DLAs. We discuss the detection of $\cto$* 
in metal-poor DLAs and its implications in Section 6. Lastly, we summarise our results and present the conclusions in Section 7.  
%
%
\section{Observations and data reduction}  
\label{sec_obs}  
The spectra of the objects studied in this project have been obtained with European Southern Observatory's 
(ESO) Ultraviolet and Visual Echelle Spectrograph (UVES) on the Very Large Telescope (VLT) [Programme ID:
086.A-0204 and 383.A-0272, PI: P. Petitjean]. For two objects, J0953$-$0504 and J0035$-$0918, we also use the Keck 
High Resolution Echelle Spectrometer (HIRES) spectra that are available in the online archive\footnote
{http://nexsci.caltech.edu/archives/koa}. One of the objects (J0953$-$0504) was selected from the sample of DLAs gathered
during the H$_{2}$ molecules survey \citep[see][]{not08}, by checking by eye those DLAs with reported [Fe/H] $\le$ $-$2.0. 
The other systems were selected from the Sloan Digital Sky Survey (SDSS) DR7 DLA catalogue \citep{not09b}, based on the 
weakness/absence of strong metal absorption lines in DLAs with large neutral hydrogen column densities. The targets were 
chosen after checking by eye those DLAs without detectable metal lines or very weak $\cto$ lines (which always remain visible 
in careful eye-checking). This is a justified method of selecting metal-poor DLAs for further study, since the strongest 
metal lines should remain undetected at the low resolution and signal-to-noise ratio (S/N) of the SDSS spectra, for the system 
to be a metal-poor DLA. However, the lack of detection does not guarantee low metallicity since it could be a consequence of 
lines with very low velocity width ($\Delta$v $\lesssim$ 10 km s$^{-1}$) being washed out by low resolution. Note that from the
correlation found by \citet{led06} we expect systems with narrow metal lines to have low metallicity. In any case, to get 
accurate estimate of the metallicities, we need to use the high-resolution spectra. \\
\indent In Table \ref{tab:obs}, we present details of the observations. All the UVES spectra were reduced using the 
Common Pipeline Library (CPL) data reduction pipeline using an optimal extraction method. All the spectra, after 
applying barycentric correction, were brought to their vacuum values using the formula given in \citet{edl66}. For 
the co-addition, we interpolated the individual spectra and their errors to a common wavelength array, and then computed
the weighted mean using the weights estimated from the error in each pixel. In the case of Keck/HIRES spectra, we use the 
pipeline calibrated data available in the Keck archive. The wavelength range covered, spectral resolution and average S/N
for each QSO are also given in Table \ref{tab:obs}. In total, we have 5 DLAs and 1 sub-DLA along the four QSO sightlines 
considered here. 
\begin{table*} 
\caption{Details of observations}
\centering
\begin{tabular}{cccccccc}
\hline
QSO                   & z$_{em}$ & z$_{abs}$ & Telescope/ & Wavelength          & Resolution    & S/N$^{b}$ & Integration                    \\
                      &          &           & instrument & range ($\AA$)$^{a}$ & (km s$^{-1}$) &           & time (s)                       \\
\hline
J003501.88$-$091817.6 & 2.413    & 2.34006   & VLT/UVES   & 3760 $-$ 9460       & 6.0           & 13        & 2$\times$3000                  \\
                      &          &           & KECK/HIRES & 3500 $-$ 6000       & 7.0           & 16        & 6$\times$2700                  \\
J023408.97$-$075107.6 & 2.540    & 2.31815   & VLT/UVES   & 3760 $-$ 9460       & 6.0           & 19        & 5$\times$3000                  \\
J095355.69$-$050418.5 & 4.369    & 4.20287   & VLT/UVES   & 4780 $-$ 6810       & 6.0           & 28        & 4$\times$7740                  \\
                      &          &           & KECK/HIRES & 6030 $-$ 8390       & 7.0           & 12        & 1$\times$7200, 1$\times$9000   \\
J100428.43$+$001825.6 & 3.045    & 2.53970   & VLT/UVES   & 4160 $-$ 6210       & 6.0           & 26        & 6$\times$3004                  \\
                      &          & 2.68537   &            &                     &               &           &                                \\
                      &          & 2.74575   &            &                     &               &           &                                \\
\hline
\end{tabular}
\label{tab:obs}
\begin{flushleft}
$^{a}$ With some wavelength gaps. \\
$^{b}$ S/N per pixel measured at $\sim$ 5000 $\AA$ (or $\sim$ 7000 $\AA$ for J0953-0504 HIRES spectrum).
\end{flushleft}
\end{table*}
%
%
\section{Data analysis}
\label{sec_analysis}
The metal absorption lines of the DLAs are modeled by Voigt profile using the VPFIT software package 
(version 9.5)\footnote{VPFIT is available from http://www.ast.cam.ac.uk/$\sim$rfc/vpfit.html}. VPFIT employs 
$\chi^{2}$ minimization to simultaneously fit Voigt profiles to a set of absorption lines, governed by three 
free parameters: (1) absorption redshift (z$_{abs}$); (2) Doppler parameter ($b$ in km s$^{-1}$); and (3) column 
density (\textit{N}). The number of components to fit was initially decided by the profiles of the unsaturated 
metal transitions in the system. We then tried to obtain a good fit by maintaining the $\chi^{2}_{red}$ 
close to 1.0, and the errors on the fitted parameters reasonable. We assumed that all the neutral and first ions 
(e.g. $\cto$, $\non$, $\oon$, $\sit$, $\sto$, $\fet$) are kinematically associated with the same gas cloud. Hence, 
the redshift and $b$ parameter for each absorption component are tied to be the same for each of the ions, i.e. 
basically considering only the turbulent component of the broadening. However, in cases where we observe that the metal
lines are very narrow, we leave both the turbulent velocity $b_{\rm turb}$ and the temperature T as free variables during 
the fitting procedure, to check if there is any significant contribution from the thermal broadening. VPFIT calculates the 
errors on each of the fitted parameters, and the errors in ion column densities quoted here are those provided by VPFIT. 
The neutral hydrogen column densities of the DLAs were determined by fitting the damping wings of the Ly$\alpha$ line 
(which are very sensitive to \textit{N}($\hon$)). The error in \textit{N}($\hon$) measurement was estimated by trying 
different continua near the Ly$\alpha$ line profile. For a given continuum, the statistical fitting error from VPFIT is 
small ($\sim$ 0.03 dex) and the error is dominated by continuum placement uncertainties.\\
\indent The abundances of elements were deduced by assuming that each element resides in a single dominant
ionization stage in the neutral gas. This assumption is valid in DLAs as they are self-shielded from metagalactic 
ionizing radiation as well as local radiation fields (for E $>$ 13.6 eV), due to the high \textit{N}($\hon$). 
While absorption of metals are seen in several components, we measure $\hon$ as one component. Hence, abundance of 
an element X is determined by taking the ratio of the total column density (sum of column densities in all detected 
individual components) of its dominant ion to that of $\hon$, and referring it to the solar scale as,
[X/H] $\equiv$ log (\textit{N}(X)/\textit{N}($\hon$)) $-$ log(\textit{N}(X)/\textit{N}($\hon$))$_{\odot}$. 
The \citet{asp09} solar scale has been used. Ionization corrections are known to be small ($\lesssim$ 0.1 dex) for the 
typical neutral hydrogen column densities expected for DLAs \citep[see][]{petj92,vla01,per07,cookb11}. In the present 
study, we did not apply any ionization correction to the abundance measurements. 
%
%
\section{Individual Systems}  
\label{ind_sys}
%
%
\subsection{z$_{abs}$ = 2.34006 towards J0035$-$0918} 
\label{sec_J0035}  
\begin{figure*}
\centering \includegraphics[width=1.0\textwidth]{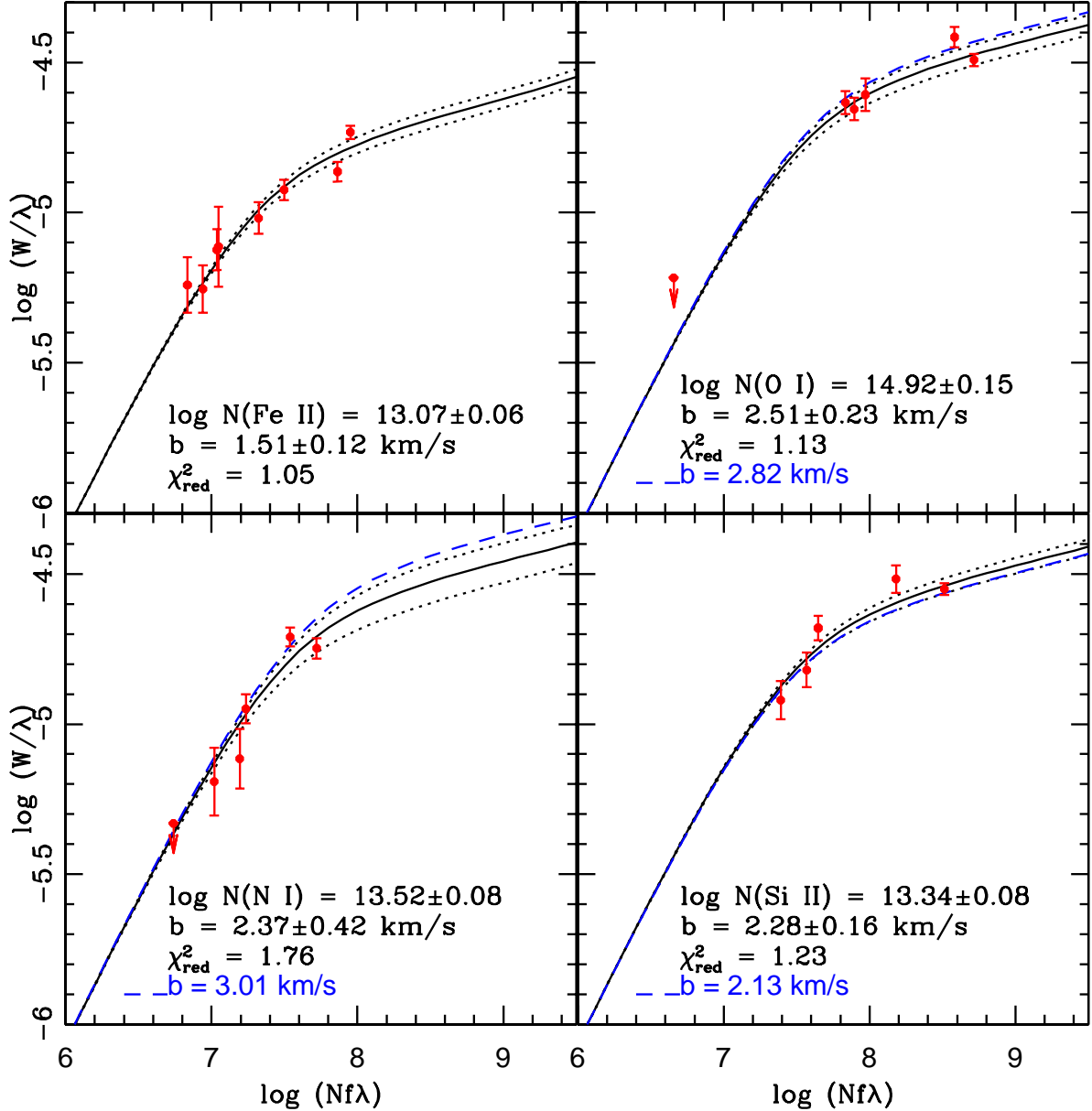}
\caption{Curves of growth for the ions detected in the DLA at z$_{abs}$ = 2.34006 towards J0035$-$0918. The black solid curves 
are obtained using the $b$ estimated from our IDL code, and the dotted lines show the 1$\sigma$ error range. The blue dashed curves are 
obtained from $b$($\fet$) under the condition of pure thermal broadening. The red points mark the observed rest-frame equivalent widths 
and the column densities calculated by our IDL code.}
\label{fig:cog}
\end{figure*}

\begin{figure}
\centering \includegraphics[width=0.4\textwidth]{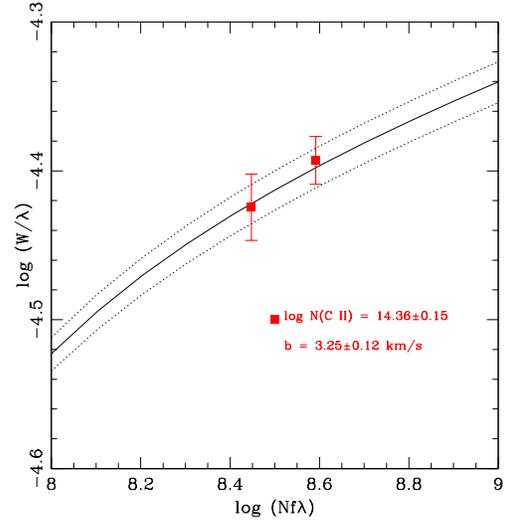}
\caption{Curve of growth for $\cto$ in the DLA at z$_{abs}$ = 2.34006 towards J0035$-$0918, obtained from $b$($\fet$). The dotted lines 
show the 1$\sigma$ error range. The red points mark the rest-frame equivalent widths of the two $\cto$ transitions. The $b$ value used 
is from $b$($\fet$) under assumption of pure thermal broadening.}
\label{fig:carboncog}
\end{figure}

\begin{figure*}
\centering
\includegraphics[width=0.8\textwidth, angle=90]{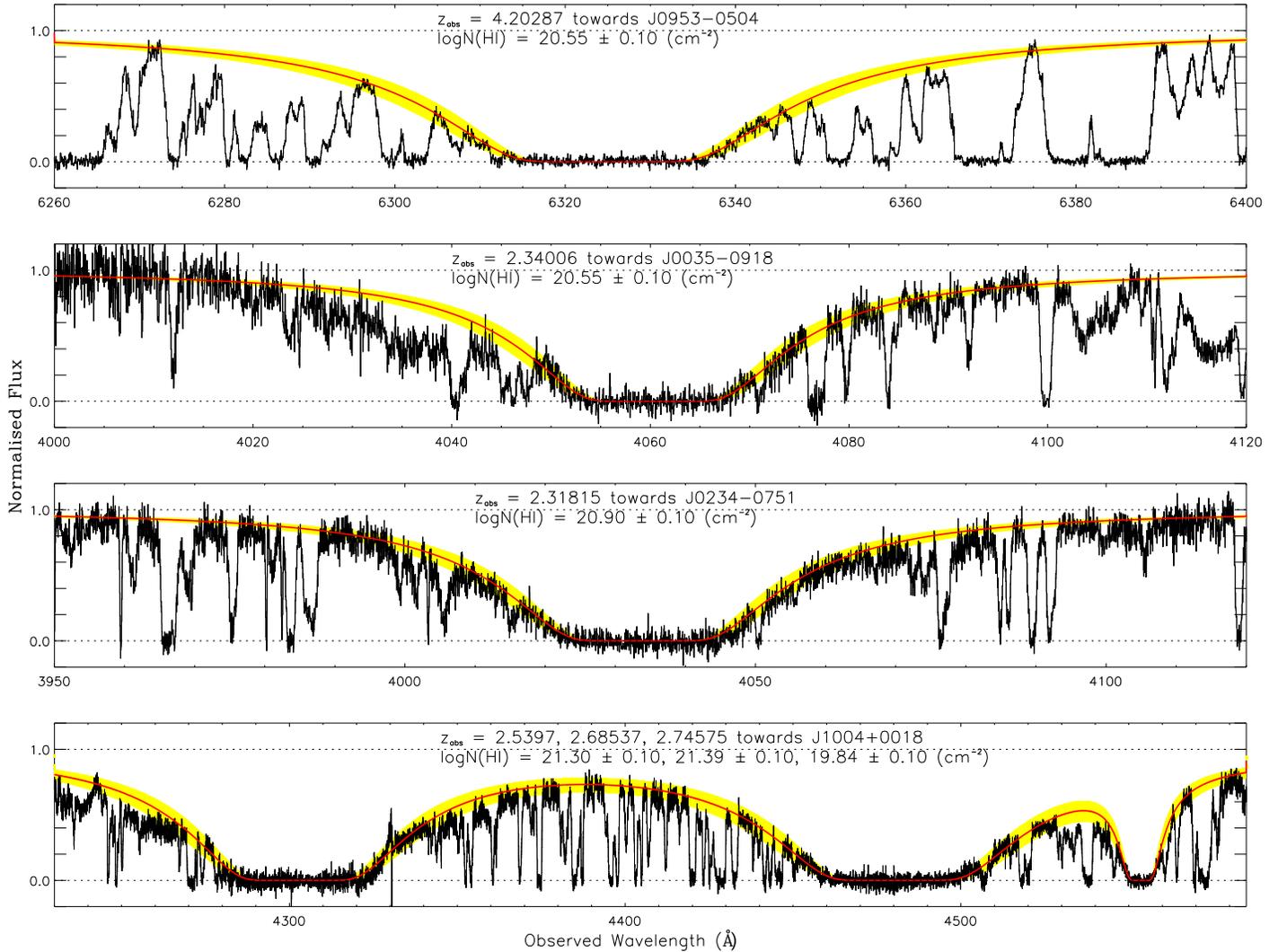}
\caption{The normalized DLA profiles overplotted with the model fit to the damped Ly$\alpha$ absorption lines in red. 
The yellow shaded regions show the 1$\sigma$ error in the fits. In each panel, QSO name, absorption redshift and the 
measured log~\textit{N}($\hon$) are also quoted.}
\label{fig:dlas}
\end{figure*}

\begin{figure*}
\centering
\includegraphics[width=0.6\textwidth, angle=90]{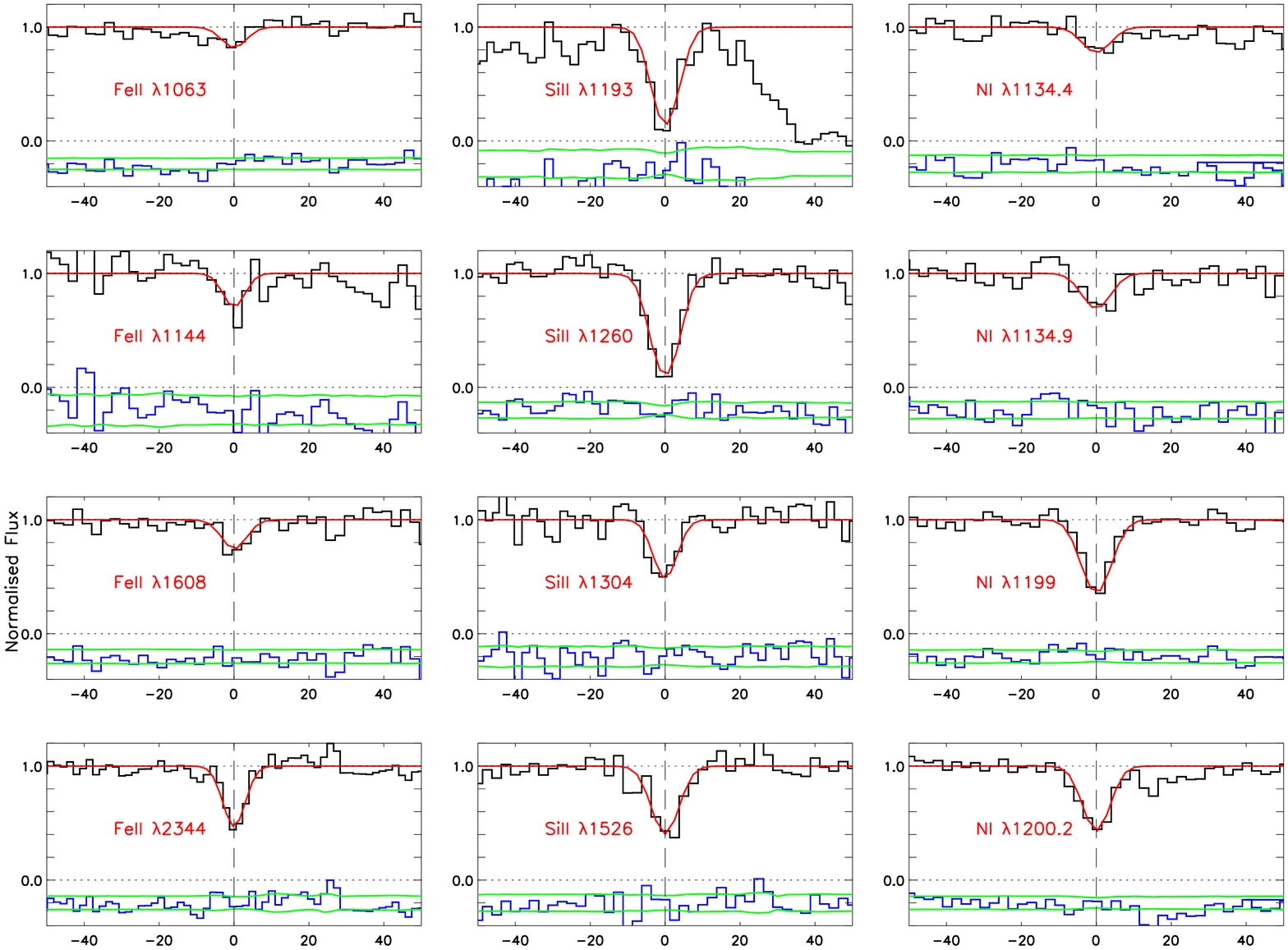}
\includegraphics[width=0.6\textwidth, angle=90]{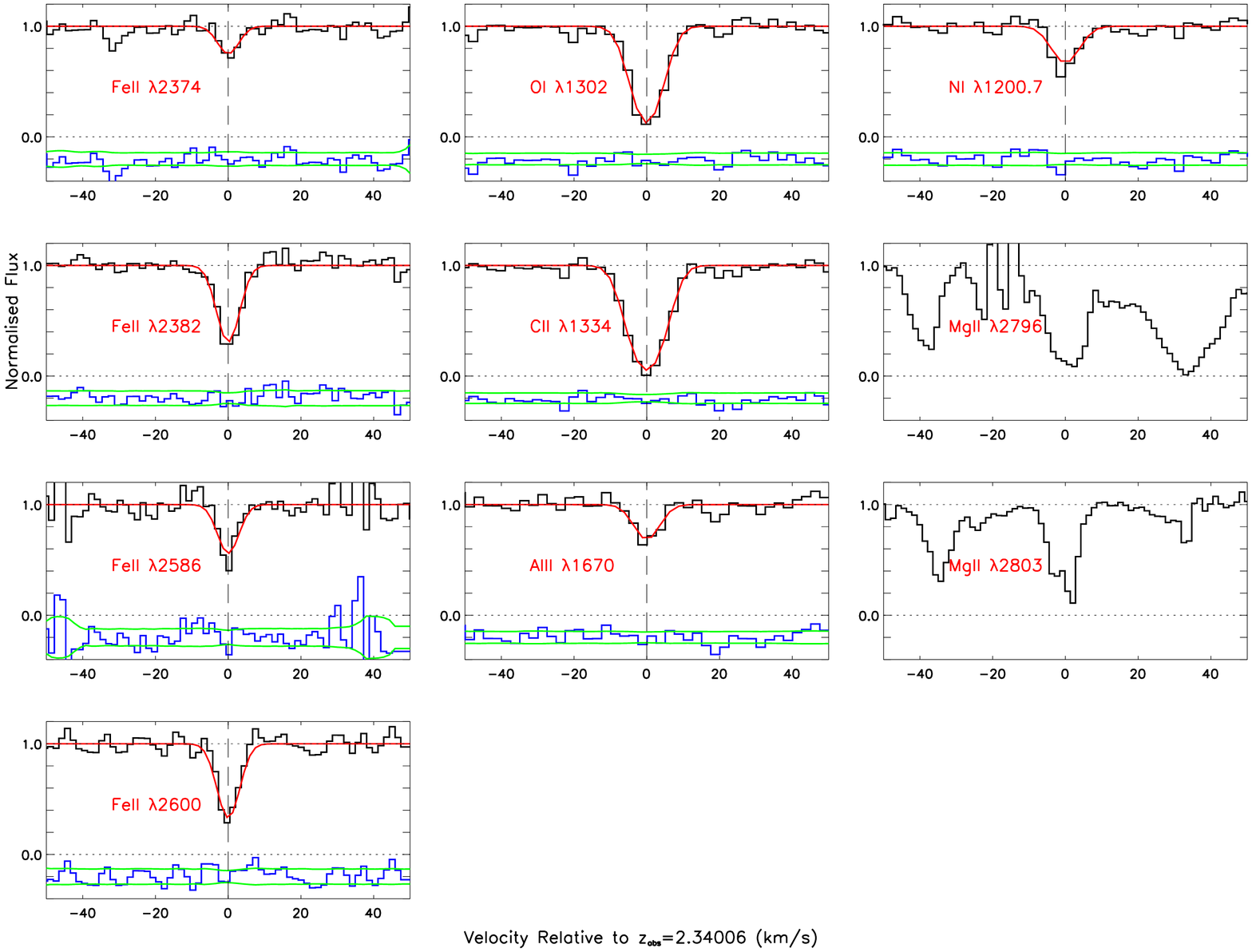}
\caption{A selection of metal lines associated with the DLA at z$_{abs}$ = 2.34006 towards J0035$-$0918. 
Best-fitting Voigt profiles are overplotted in red. The dashed vertical lines show the component positions. 
The errors in flux and residuals from the fit are shown at the bottom in green and blue respectively.
No fit is performed for the $\mgt$ lines as these wavelength ranges are contaminated by atmospheric absorption.}
\label{fig:ionsj0035}
\end{figure*}

\begin{figure*}
\centering
\includegraphics[width=0.3\textwidth, angle=90]{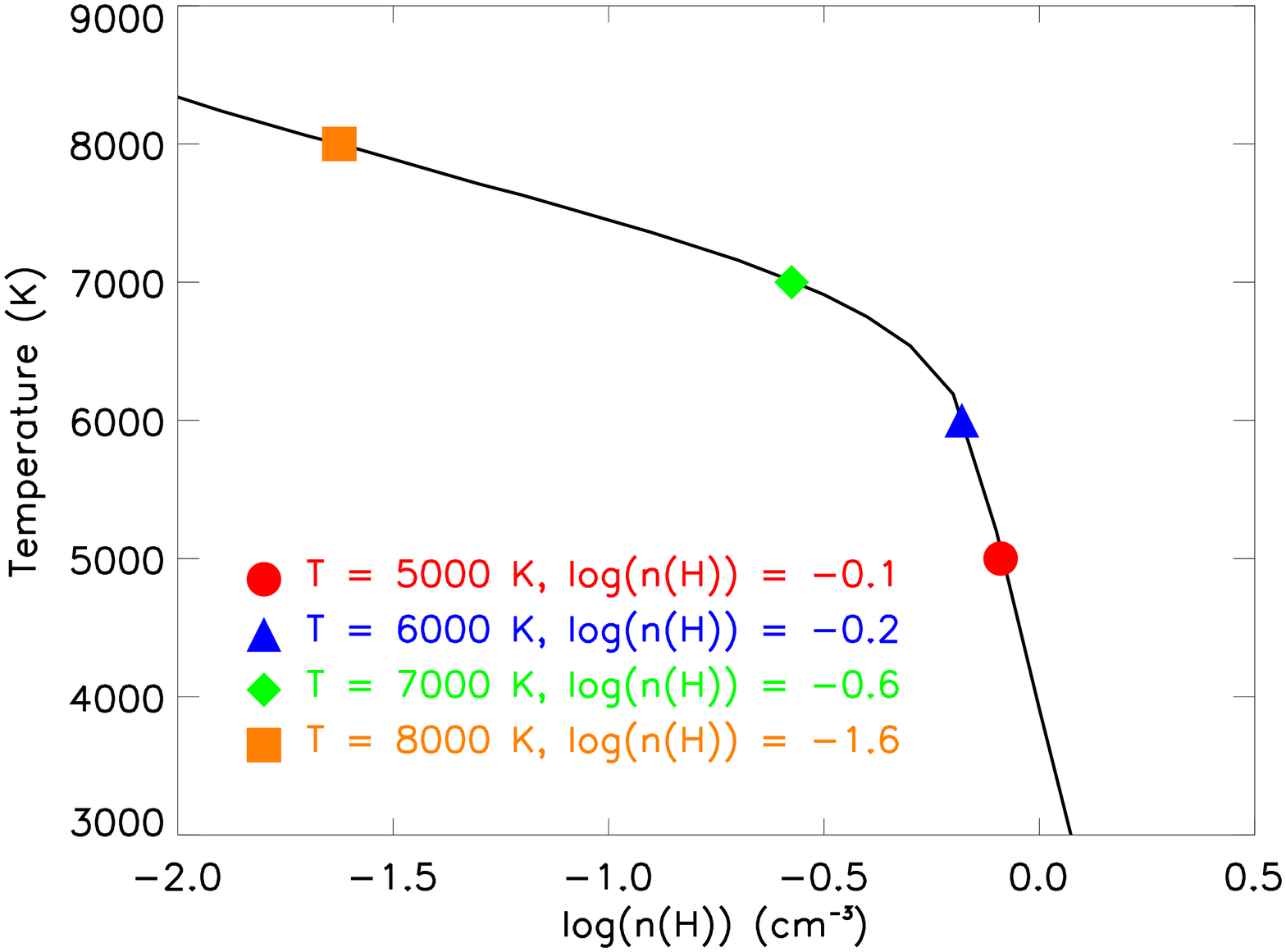}
\includegraphics[width=0.3\textwidth, angle=90]{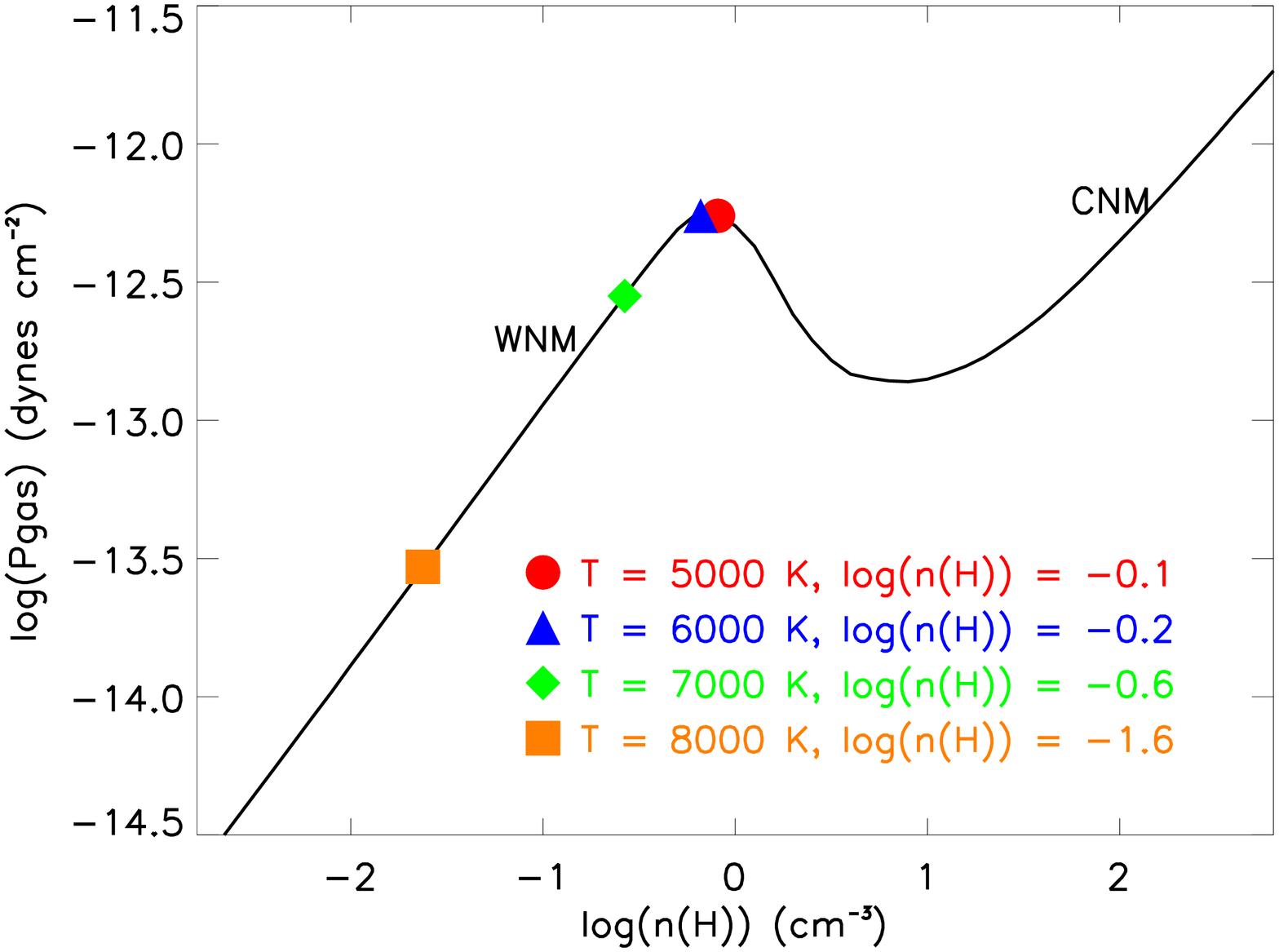}
\caption{Results of photoionization calculations using \textit{N}($\hon$) and metallicity similar to the observed values for the z$_{abs}$ = 2.34006 
system towards J0035$-$0918. For the ionizing field similar to that of UV background at z $\sim$ z$_{abs}$, the left panel shows the gas temperature as
a function of hydrogen density. The corresponding phase diagram is shown in the right panel.}
\label{fig:j0035cloudy}
\end{figure*}

We identified this DLA as metal-poor by the lack of strong metal lines in its SDSS spectrum. It was subsequently observed by 
UVES on VLT on 28th and 29th of December, 2010. Meanwhile, this VMP DLA has also been studied by \citet{cooka11} using Keck/HIRES 
spectrum, and they found it to be the most carbon-enhanced ([C/Fe] = 1.53) metal poor DLA ([Fe/H] $\simeq$ $-$3) detected till date. 
The wavelength range covered by our UVES spectrum (3760 $-$ 9640 $\AA$) differs from that of the spectrum used by \citet{cooka11} 
(3100 $-$ 6000 $\AA$). Our spectrum covers more $\fet$ transitions than that used by \citet{cooka11}, enabling us to get a better constrained 
measurement of $\fet$ abundance and contribution of thermal broadening to the line profiles. We also use the archival Keck/HIRES spectrum 
(3500 $-$ 6000 $\AA$) to get measurements of a few metal lines not covered by our spectra. Unfortunately, our UVES spectrum covers only one 
$\cto$ line ($\lambda$1334) and $\oon$ line ($\lambda$1302), both of which are nearly saturated. As we do not have access to the part of the 
Keck spectrum covering the other $\oon$ and $\cto$ transitions, as well as a few other transitions of interest, we adopt the measurements of 
equivalent widths of these lines from \citet{cooka11}. In Table \ref{tab:eqw}, we give details of the metal lines and their equivalent widths. \\
\indent Since all the metal lines show single component structure and are unblended, we first carry out a curve of growth analysis for this system. 
We have written an Interactive Data Language (IDL) code using the MPFIT routine \citep{mark09}, to estimate the $b$ parameter and column density 
which best fit the observed equivalent widths of different lines of an ion. The code treats the quantity W$_{0}$/$\lambda$\textit{N} as function of 
\textit{f}$\lambda$ (W$_{0}$: rest-frame equivalent width, $\lambda$: wavelength, \textit{f}: oscillator strength), and estimates the 
combination of the parameters \textit{N} and $b$ which gives the best-fitting curve of growth. Using this we first calculate the $b$ and \textit{N} 
for $\fet$, for which we can get a robust estimate, since we have eight transitions covering a wide range of \textit{f}$\lambda$ ($>$ 1 dex) values. 
We find that we get $b$($\fet$) = 1.51 $\pm$ 0.12 km s$^{-1}$ and log[\textit{N}($\fet$)(cm$^{-2}$)] = 13.07 $\pm$ 0.06. We also fit all the $\fet$ 
lines covered in our UVES spectrum using VPFIT and get similar values [$b$($\fet$) = 1.63 $\pm$ 0.11 km s$^{-1}$ and log[\textit{N}($\fet$)(cm$^{-2}$)] 
= 13.10 $\pm$ 0.05]. Then we go on similarly to calculate the $b$ and \textit{N} values for $\oon$, $\sit$ and $\non$. We plot the curves of growth using 
the best estimated $b$ parameters in Fig. \ref{fig:cog}. From this figure we can clearly see that different metal lines require different $b$ values, 
indicating that the thermal contribution to the $b$ parameter is significant. We also show in the same plot, the expected curves of growth for $\oon$, $\non$ 
and $\sit$ with $b$ parameters estimated from $b$($\fet$) assuming only thermal broadening. These seem to almost match within errors with the curves obtained 
using the $b$ from our IDL code, showing that thermal broadening is dominant. It is difficult to get a proper estimate of \textit{N}($\cto$) even from the Keck 
spectrum used by \citet{cooka11}, since it covers two transitions of $\cto$ ($\lambda$1036 and $\lambda$1334), with similar values of \textit{f}$\lambda$, which 
fall on the flat part of the curve of growth. Here, we calculate $b$($\cto$) = 3.25 $\pm$ 0.12 km s$^{-1}$ from $b$($\fet$) assuming only thermal broadening. 
From this we estimate log[\textit{N}($\cto$)(cm$^{-2}$)] = 14.36 $\pm$ 0.15, by requiring that the observed equivalent widths of the two transitions are 
satisfied by the resultant curve of growth (see Fig. \ref{fig:carboncog}). \\
\indent Additionally, to check the results from the curve of growth method, we use VPFIT to fit the metal lines covered in our UVES spectrum. 
Firstly, we find that if we constrain the $b$ parameter to be the same for all the ions (i.e. neglect the thermal broadening), a single component 
cloud model with $b$ = 2.00 $\pm$ 0.05 km s$^{-1}$ fits the observed line profiles. Next, we fit the metal lines using $b_{\rm turb}$ and T as 
independent variables for two ions of different mass ($\fet$ and $\non$) at the same redshift simultaneously. The chi-square for this fit is better 
than that for the fit considering only turbulent broadening. The best-fit values obtained are, $b_{\rm turb}$ = 0.7 $\pm$ 0.5 km s$^{-1}$ and 
T = (7.6 $\pm$ 1.9) $\times$ 10$^{3}$ K. These values indicate that in this case, the line widths are mainly determined by thermal 
broadening. In Table \ref{tab:coldenj0035}, we provide comparison of the ion column densities obtained from the two different VPFIT fits and also 
those obtained from the curve of growth technique. The large errors in $b_{\rm turb}$, T, \textit{N}($\oon$) and \textit{N}($\cto$) that we get from 
VPFIT are likely to be due to the low S/N of the UVES spectrum and the fact that all the $\oon$ and $\cto$ lines are not available to us for fitting. 
The values from the fit which considers both thermal and turbulent components of the Doppler parameter can be seen to agree well with those obtained 
from the curve of growth analysis. We find that the greatest variation in column densities, obtained by the above two methods and from the turbulence-only 
fit, occurs for $\oon$ ($\sim$ 1 dex) and $\cto$ ($\sim$ 2 dex). This is expected as both our $\oon$ and $\cto$ lines fall on the saturated part of the 
curve of growth, where the column density depends critically on the assumed broadening mechanism. The most important result from the comparison given in 
Table \ref{tab:coldenj0035}, is that the enhancement of carbon over iron (and also over oxygen) reduces drastically if we consider thermal broadening as 
the dominant broadening mechanism instead of turbulence. Using the curve of growth results we get [C/Fe] = 0.36 $\pm$ 0.16, and from the best-fitting model 
using VPFIT we obtain [C/Fe] = 0.45 $\pm$ 0.19. The C enhancement is then $\sim$ 3 times instead of $\sim$ 30 times as reported by \citet{cooka11}. Indeed, 
\citet{car12} also note the same and they report [C/Fe] = 0.51 $\pm$ 0.10 for a thermal fit to the system using VPFIT, without giving any details. They obtain 
T = (7.66 $\pm$ 0.57) $\times$ 10$^{3}$ K, similar to our values, however they get a smaller error. Here we have carried out a detailed analysis 
using both curve of growth and VPFIT, using more transitions of $\fet$. We come to the conclusion that the metal lines are most likely to be thermally broadened, 
and while the [C/Fe] ratio may be slightly above the range seen in metal-poor DLAs ($-$0.1 to 0.4) \citep{cookb11}, C abundance is not more than $\sim$ 3 
times that of Fe. \\
\indent A value of log[\textit{N}($\hon$)(cm$^{-2}$)] = 20.55 $\pm$ 0.10 is obtained for this DLA using our UVES spectrum, which is consistent with
the value of \citet{cooka11}, and also \citet{jor13} (see Fig. \ref{fig:dlas} for the fit to the DLA profile). We have derived the 3$\sigma$ limiting 
rest-frame equivalent width for $\sto$ using the strongest undetected transition ($\lambda$1259), over the Full-Width-at-Half-Maximum (FWHM) obtained 
from our best-fitting VPFIT model. We use it to calculate the upper limit to the $\sto$ column density in the optically thin limit approximation (\textit{N} 
= 1.13 $\times$ 10$^{20}\ldotp$W$_{0}$/$\lambda^{2}$\textit{f}). Our spectrum also covers the $\mgt$ doublet $\lambda\lambda$ 2796, 2803 (see Fig. 
\ref{fig:ionsj0035} for a selection of the metal line profiles). However, these lines are affected by the atmospheric absorption lines, so we can 
only estimate an upper limit to the column density as, log[\textit{N}($\mgt$)(cm$^{-2}$)] $\le$ 13.44. We have modified the continua near 
the $\sit$ $\lambda$1193 and $\lambda$1260 profiles for fitting purposes. In Table \ref{tab:colden}, we present the measured column densities of 
some selected ions and the abundance measurements are given in Table \ref{tab:met}. Note that the fits to the metal lines and their column 
densities and metallicities that we present are those from the thermal fit using VPFIT. \\
\indent From our curve of growth analysis for $\fet$, we get a temperature of 7600 $\pm$ 1200 K, assuming only thermal broadening.
If we consider $\fet$ and $\oon$ as reference ions to solve for both temperature and $b_{\rm turb}$, we obtain T $\sim$ 5500 K and $b_{\rm turb}$ 
$\sim$ 0.80 km s$^{-1}$. These values are consistent with what is estimated by VPFIT. Hence, we can conclude that for this system, the $b_{\rm turb}$ is
$<$ 1.0 km s$^{-1}$ and temperature lies in the range 5000 $\sim$ 8000 K. The temperature range is comparable to that expected in the Warm Neutral 
Medium (WNM) (5000 $\sim$ 10000 K) \citep{wolf03a}. The high temperature is also consistent with the non-detection of $\cto$* in this system, 
which is most likely to arise in the Cold Neutral Medium (CNM) \citep{wolf03a,wolf03b}. From the 3$\sigma$ limiting rest-frame equivalent width 
of the $\cto$* $\lambda$ 1335.71 line, we estimate log[\textit{N}($\cto$*)(cm$^{-2}$)] $\le$ 12.30. Since we have an estimate of temperature for this 
system, we model it using the photoionization software CLOUDY \citep[version 07.02.02; developed by][]{fer98}, as a plane-parallel slab of gas exposed 
to the metagalactic UV (Ultraviolet) background radiation field given by \citet{hm01}. In the left panel of Fig. \ref{fig:j0035cloudy}, we plot the 
temperatures estimated for different values of the neutral hydrogen density. From this figure, we can see that for temperatures ranging from 5000 
to 8000 K, the density ranges from 0.8 to 0.02 cm$^{-3}$, which gives length scale of the cloud between $\sim$ 4 kpc and $\sim$ 100 pc. The column
density of $\cto$* predicted by the CLOUDY model is consistent with the upper limit estimated. We also plot the phase diagram (i.e. gas pressure vs. 
neutral hydrogen density) predicted by the CLOUDY model in the right panel of Fig. \ref{fig:j0035cloudy}. It is clear from this figure that the observed
temperature range and the inferred density range correspond to the WNM phase. Temperature estimate based on the line widths in DLAs is difficult due to 
blending of the velocity components, and measurements based on narrow absorption components is likely to correspond to the CNM. We note that this DLA 
is among the few DLAs where temperature estimation is possible and the temperature obtained from the inferred line widths is as expected in the WNM 
\citep[see][]{car12,not12}.
\begin{table} 
\caption{Equivalent widths of metal lines in the DLA at z$_{abs}$ = 2.34006 towards J0035$-$0918}
\centering
\begin{tabular}{cccccc}
\hline
Ion    & Wavelength$^{a}$ & \textit{f}$^{a}$ & \textit{W$_{0}$}$^{b}$ & \textit{$\delta$W$_{0}$}$^{c}$ & Ref. \\     
       & ($\AA$)          &                  &  ($\AA$)               & ($\AA$)                        &      \\
\hline
$\cto$ & 1036.3367        & 0.1180           & 0.0390                 & 0.0020                         & 3    \\
$\cto$ & 1334.5323        & 0.1278           & 0.0540                 & 0.0020                         & 3    \\
$\non$ & 1134.1653        & 0.0146           & $<$0.0053              & ---                            & 3    \\
$\non$ & 1134.4149        & 0.0278           & 0.0073                 & 0.0019                         & 2    \\
$\non$ & 1134.9803        & 0.0416           & 0.0087                 & 0.0020                         & 2    \\
$\non$ & 1199.5496        & 0.1320           & 0.0215                 & 0.0017                         & 2    \\
$\non$ & 1200.2233        & 0.0869           & 0.0234                 & 0.0017                         & 2    \\
$\non$ & 1200.7098        & 0.0432           & 0.0135                 & 0.0015                         & 2    \\
$\oon$ & 971.7382         & 0.0116           & 0.0240                 & 0.0030                         & 3    \\
$\oon$ & 988.5778         & 0.000553         & $<$0.0060              & ---                            & 3    \\
$\oon$ & 988.6549         & 0.0083           & 0.0230                 & 0.0020                         & 3    \\
$\oon$ & 988.7734         & 0.0465           & 0.0380                 & 0.0030                         & 3    \\
$\oon$ & 1039.2304        & 0.00907          & 0.0230                 & 0.0020                         & 3    \\
$\oon$ & 1302.1685        & 0.0480           & 0.0420                 & 0.0020                         & 3    \\
$\alt$ & 1670.7886        & 1.7400           & 0.0134                 & 0.0022                         & 2    \\
$\sit$ & 989.8731         & 0.1710           & 0.0150                 & 0.0020                         & 3    \\
$\sit$ & 1193.2897        & 0.5820           & 0.0363                 & 0.0038                         & 1    \\
$\sit$ & 1260.4221        & 1.1800           & 0.0355                 & 0.0016                         & 1    \\
$\sit$ & 1304.3702        & 0.0863           & 0.0157                 & 0.0023                         & 1    \\
$\sit$ & 1526.7070        & 0.1330           & 0.0319                 & 0.0003                         & 2    \\
$\sto$ & 1259.5180        & 0.0166           & $<$0.0030              & ---                            & 1    \\
$\fet$ & 1063.1764        & 0.0547           & 0.0061                 & 0.0013                         & 2    \\
$\fet$ & 1144.9379        & 0.0830           & 0.0088                 & 0.0027                         & 1    \\
$\fet$ & 1608.4509        & 0.0577           & 0.0121                 & 0.0019                         & 2    \\
$\fet$ & 2344.2130        & 0.1140           & 0.0279                 & 0.0022                         & 1    \\
$\fet$ & 2374.4603        & 0.0313           & 0.0132                 & 0.0024                         & 1    \\
$\fet$ & 2382.7642        & 0.3200           & 0.0441                 & 0.0023                         & 1    \\
$\fet$ & 2586.6496        & 0.0691           & 0.0248                 & 0.0030                         & 1    \\
$\fet$ & 2600.1725        & 0.2390           & 0.0356                 & 0.0027                         & 1    \\
\hline
\end{tabular}
\label{tab:eqw}
\begin{flushleft}
$^{a}$ Laboratory wavelengths and oscillator strengths (\textit{f}) from \citet{mor03}. \\
$^{b}$ Rest-frame equivalent width, 
$^{c}$ Error in rest-frame equivalent width. \\
$^{1}$ VLT/UVES spectrum, 
$^{2}$ Keck/HIRES archival spectrum, 
$^{3}$ \citet{cooka11}. 
\end{flushleft}
\end{table}
\begin{table} 
\caption{Column densities of ions in the DLA at z$_{abs}$ = 2.34006 towards J0035$-$0918}
\centering
\begin{tabular}{cccc}
\hline
Ion    &               & log~\textit{N} (cm$^{-2}$) &            \\
       & VPFIT 1$^{a}$ & VPFIT 2$^{b}$              & COG $^{c}$ \\ 
\hline
$\cto$ & 16.18 (0.11)  & 14.45 (0.19)  & 14.36 (0.15) \\
$\non$ & 13.60 (0.04)  & 13.48 (0.03)  & 13.52 (0.08) \\
$\oon$ & 15.68 (0.18)  & 14.55 (0.14)  & 14.92 (0.15) \\
$\sit$ & 13.50 (0.07)  & 13.37 (0.06)  & 13.34 (0.08) \\
$\fet$ & 13.01 (0.03)  & 13.07 (0.04)  & 13.07 (0.06) \\
\hline
\end{tabular}
\label{tab:coldenj0035}
\begin{flushleft}
$^{a}$ $b$ = 2.00 $\pm$ 0.05 km s$^{-1}$ ; $\chi^{2}$/d.o.f. = 1010/859 \\
$^{b}$ $b_{\rm turb}$ = 0.89 $\pm$ 0.41 km s$^{-1}$ and T = (7.3 $\pm$ 1.7) $\times$ 10$^{3}$ K ; $\chi^{2}$/d.o.f. = 999/858 \\
$^{c}$ Curve of growth analysis 
\end{flushleft}
\end{table}
%
%
\subsection{z$_{abs}$ = 2.31815 towards J0234$-$0751}  
\label{sec_J0234}  
\begin{figure*}
\centering
\includegraphics[width=0.8\textwidth, angle=90]{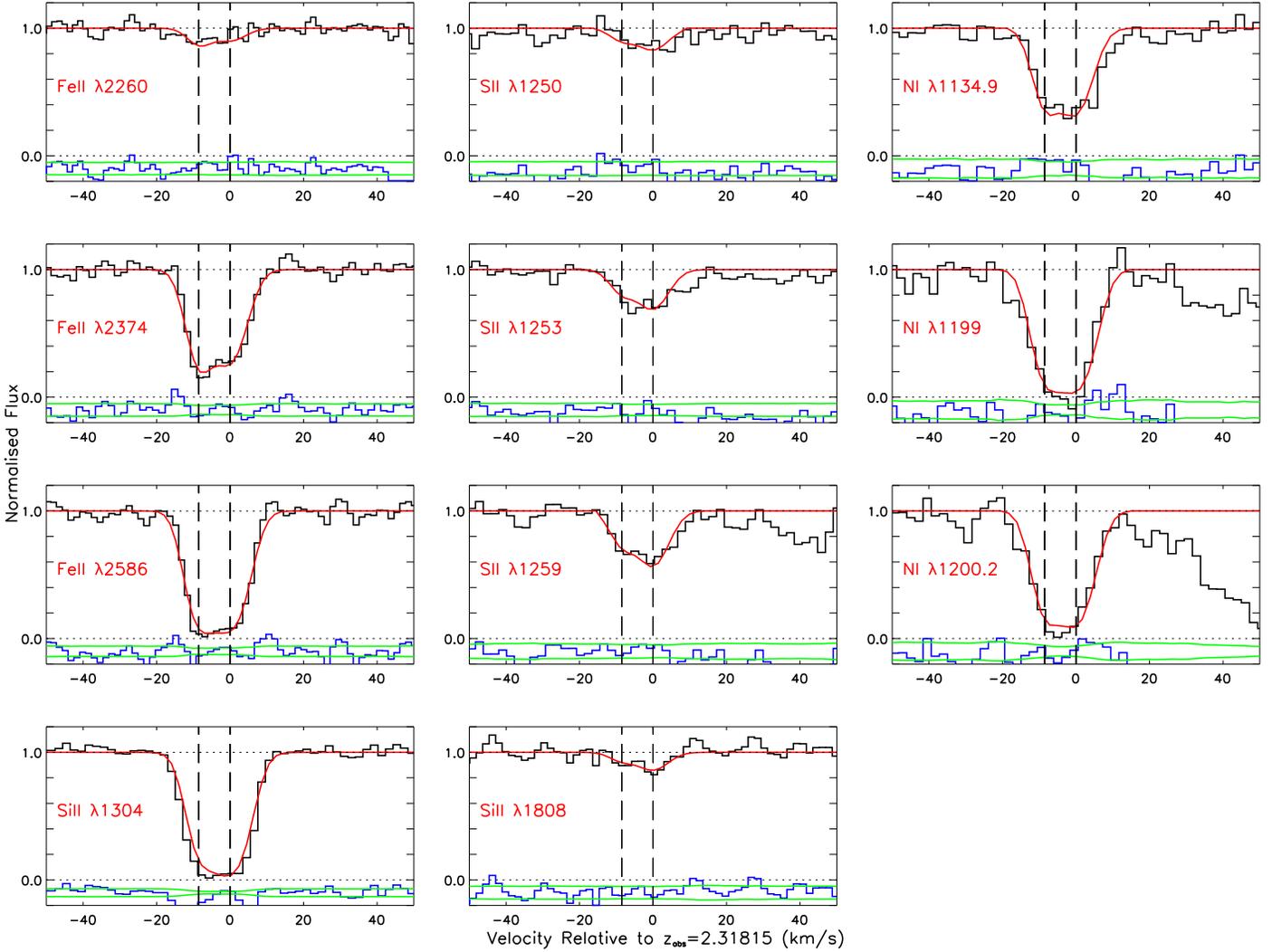}
\caption{Same as in Fig. \ref{fig:ionsj0035} for a selection of metal lines associated with the DLA at z$_{abs}$ = 2.31815 towards J0234$-$0751.}
\label{fig:ionsj0234}
\end{figure*}

This source was picked from the SDSS database based on the absence of strong metal lines in the spectrum. We obtain a value of 
log[\textit{N}($\hon$)(cm$^{-2}$)] = 20.90 $\pm$ 0.06 (see Fig. \ref{fig:dlas} for the fit to DLA profile) for this system. This 
DLA  qualifies as metal poor with [Fe/H] $\sim$ $-$2.23. The $\fet$ abundance is well-constrained by the unsaturated transitions 
of $\lambda$2260 and $\lambda$2374. However, the transitions of $\oon$ $\lambda$1302 and $\cto$ $\lambda$1334 (the only $\oon$ 
and $\cto$ lines covered by our spectrum) are both saturated. We estimate lower limits to their column densities using the optically 
thin limit approximation as given in section \ref{sec_J0035}. We notice from Table 1 of \citet{wolf08} that log[\textit{N}($\hon$)(cm$^{-2}$)] = 
20.95 $\pm$ 0.15, [M/H] = $-$2.74 $\pm$ 0.14 and [Fe/H] = $-$3.14 $\pm$ 0.04 was reported for this system without details, while 
Table 1 of \citet{jor13} gives log[\textit{N}($\hon$)(cm$^{-2}$)] = 20.85 $\pm$ 0.10, [Si/H] = $-$2.46 $\pm$ 0.10 and [Fe/H] = $-$2.56 $\pm$ 0.04 
for this system, from medium resolution spectra, without details. While our \textit{N}($\hon$) measurement matches with that of these 
two cases, we disagree with their metallicity estimates. \\
\indent A two component cloud model (with $b$ parameters of 2.71 $\pm$ 0.33 and 4.29 $\pm$ 0.38 km s$^{-1}$ separated by $\sim$ 8.5 
km s$^{-1}$) is found to fit well with the observed ion profiles (see Fig. \ref{fig:ionsj0234} for a selection of the metal profiles). 
Since the metal lines are narrow, we also did a fit using $b_{\rm turb}$ and T of two ions of different mass ($\fet$ and $\non$) as 
independent variables, in order to check whether there is any significant contribution from thermal broadening. However, the resulting 
fit was poor and the chi-square ($\chi^{2}_{red}$ = 2.0) higher than that obtained for our best-fitting model ($\chi^{2}_{red}$ 
= 1.4). Hence, we use our best-fitting result which assumes that the thermal broadening is negligible in comparison to the turbulent 
broadening. The details of selected ion column densities and abundances are provided in Table \ref{tab:colden} and Table 
\ref{tab:met} respectively. 
%
%
\subsection{z$_{abs}$ = 4.20287 towards J0953$-$0504}  
\label{sec_J0953}  
\begin{figure*}
\centering
\includegraphics[width=0.8\textwidth, angle=90]{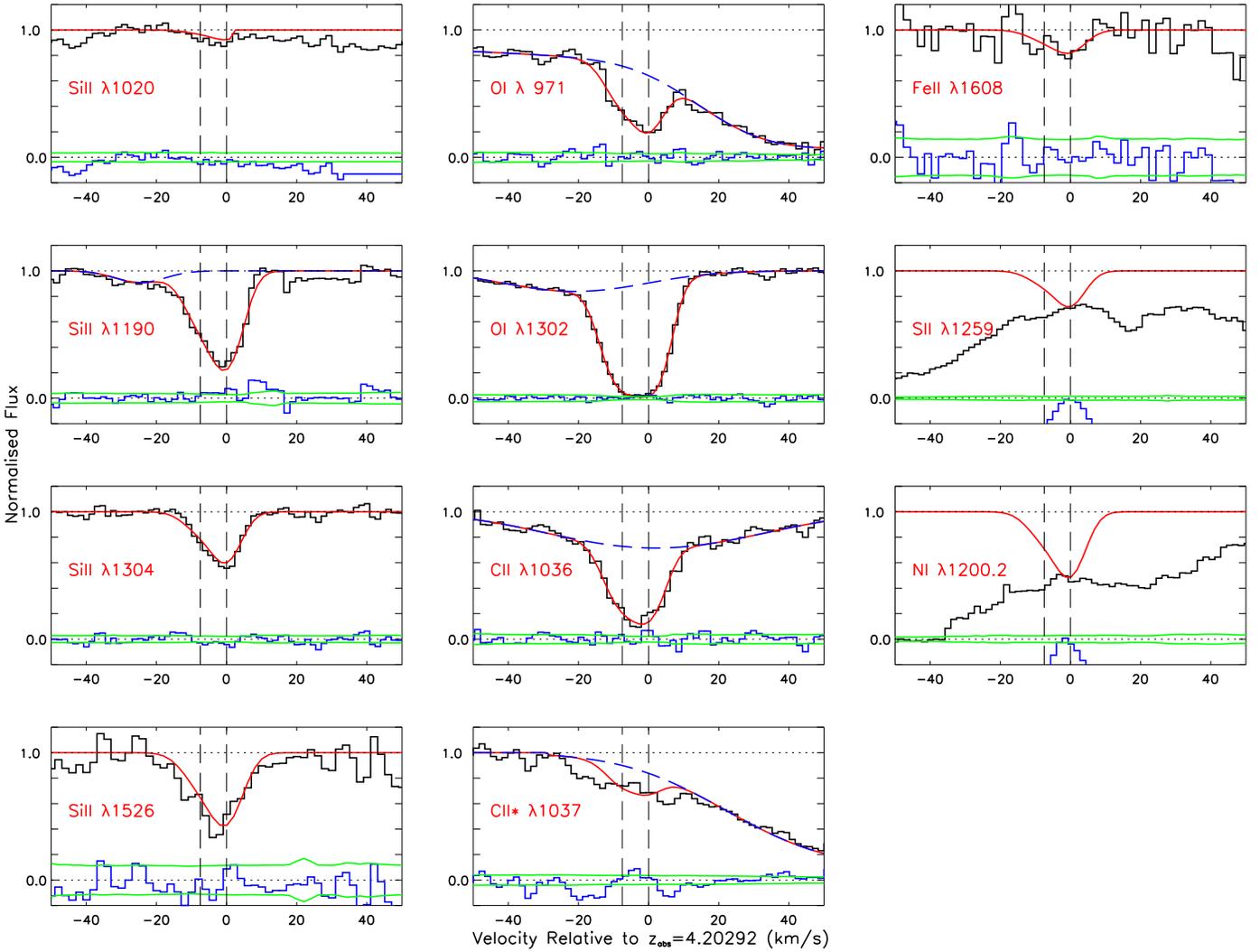}
\caption{Same as in Fig. \ref{fig:ionsj0035} for a selection of metal lines associated with 
the DLA at z$_{abs}$ = 4.20287 towards J0953$-$0504. The dashed blue lines show the Ly$\alpha$ 
profiles which blend the lines of interest.}
\label{fig:ionsj0953}
\end{figure*}

This system (also known as Q0951$-$0450) is part of a DLA sample used by \citet{not08} to search for H$_{2}$ in DLAs. For this particular 
sightline we have both VLT/UVES spectra (4780 $-$ 6810 $\AA$) and Keck/HIRES spectra (6030 $-$ 8390 $\AA$). Almost all our lines of interest 
fall in the wavelength range covered by the UVES spectra. For two transitions not covered by the UVES spectrum ($\sit$ $\lambda$1526 \& 
$\fet$ $\lambda$1608), we use the HIRES spectrum to fit the profiles. We derive log[\textit{N}($\hon$) (cm$^{-2}$)] = 20.55 $\pm$ 0.10 
for this system, which is consistent with the measurement of \citet{not08}. Fig. \ref{fig:dlas} shows the best-fitting Voigt profile overplotted 
on the Ly$\alpha$ profile. A two component cloud model (with $b$ parameters of 4.43 $\pm$ 0.28 and 5.33 $\pm$ 0.60 km s$^{-1}$ separated by $\sim$ 
7.5 km s$^{-1}$) is found to give a good fit to the observed metal line profiles (see Fig. \ref{fig:ionsj0953} for a selection of the metal 
lines). From Table \ref{tab:j0953}, we can see that the component with $b$ $\sim$ 4 km s$^{-1}$, contains about 60\% of the $\cto$ and $\oon$, 
and about 70\% of the $\sit$ and $\fet$ present in the cloud. \\
\indent All the $\fet$ lines except $\lambda$1608 are blended/undetected, and the error on $\fet$ column density is relatively large 
since only one line in a low S/N region could be used for fitting. Even allowing for the maximum error ($0.21$ dex) in the metallicity 
measurement, this system is among one of the most metal-poor DLAs, and the most metal-poor DLA at z $>$ 4 detected till date. However, 
most of the lines of interest fall in the Ly$\alpha$ forest and hence are blended, making identification and fitting of line profiles not 
straightforward. There is also another DLA along this line of sight (at z$_{abs}$ = 3.8567), whose metal lines sometimes contaminate our 
lines of interest. For the system at z$_{abs}$ = 4.20287, our spectrum covers one unsaturated line each of $\oon$ ($\lambda$971) and 
$\cto$ ($\lambda$1036) (see Fig. \ref{fig:ionsj0953}). These lines allow us to derive reliable abundance measurements of O and C respectively. 
The $\oon$ $\lambda$971 line is in the wing of a Ly$\alpha$ absorption and we fit it along with the Ly$\alpha$ line. We also overplot the best-fitting 
profile over the saturated $\oon$ $\lambda$1302 line and find it to be consistent. The $\cto$ $\lambda$1036 line is blended and we fit it by assuming 
the contamination to be a broad Ly$\alpha$. We obtain an upper limit to the $\cto$* column density by fitting the $\lambda$1037 line profile along with 
a Ly$\alpha$ profile. All the Ly$\alpha$ profiles which blend our lines of interest are shown in dotted blue lines in Fig. \ref{fig:ionsj0953}. Since 
all the $\sto$ and $\non$ profiles are blended, we estimate upper limits to their column densities by overplotting profiles, obtained using the 
parameters from the best-fitting model, on the strongest unsaturated transition profiles ($\sto$ $\lambda$1259 \& $\non$ $\lambda$1200.2), and 
adjusting the column densities till the maximum limit. We show the absorption profiles with maximum column densities in Fig. \ref{fig:ionsj0953}. 
The details of selected ion column densities are presented in Table \ref{tab:colden} and abundances are presented in Table \ref{tab:met}. \\
\indent With [Fe/H] $\simeq$ $-$3.0, this system is the only known extremely metal-poor DLA detected till date at z $>$ 4. The relative abundance 
pattern of elements in this DLA is found to be similar to that of a typical metal-poor DLA \citep[see][]{cookb11}, as shown in Fig. \ref{fig:comp}. 
We find no enhancement of C over Fe as such ([C/Fe] = $-$0.07 $\pm$ 0.19). Also, C is not enhanced with respect to O ([C/O] = $-$0.5 $\pm$ 0.03). 
So like most metal-poor DLAs, typical Population II star yields will explain the abundance pattern seen in this DLA. Oxygen is found to be enhanced 
with respect to iron, [O/Fe] = 0.43 $\pm$ 0.20. While the present measurement is consistent with [$\langle$O/Fe$\rangle$] = 0.35 $\pm$ 0.09 found in 
DLAs with $-$3 $\le$ [Fe/H] $\le$ $-$2 \citep[see][]{cookb11}, it is slightly lower than the mean [$\langle$O/Fe$\rangle$] = 0.69 $\pm$ 0.14 measured 
for three [Fe/H] $<$ $-$3 DLAs in the sample of \citet{cookb11}. Therefore, a possible trend of increasing [O/Fe] at [Fe/H] $<$ $-$3 as suggested by 
\citet{cookb11}, requires confirmation with more measurements of these metallicities. Using the upper limit on \textit{N}($\non$) obtained as described 
previously, we find that [N/O] $\le$ $-$0.27. This lies above the primary plateau (see section \ref{sec_no} for more details), and unfortunately we 
cannot come to any definite conclusions regarding the production of nitrogen in this system. \citet{wolf08} have reported log[\textit{N}($\cto$*)(cm$^{-2}$)] = 
13.37 $\pm$ 0.08 for this system, without showing the spectrum. Our spectrum covers only the $\cto$* $\lambda$1037 line. This line is blended with a 
nearby Ly$\alpha$ absorption and our fits suggest log[\textit{N}($\cto$*)(cm$^{-2}$)] $\le$ 12.95 (see Fig. \ref{fig:ionsj0953}). 
\begin{table} 
\caption{Component-wise distribution of column densities of ions in the DLA at z$_{abs}$ = 4.20287 towards J0953$-$0504}
\centering
\begin{tabular}{ccc}
\hline
Ion    &  log~\textit{N} (cm$^{-2}$)            &  log~\textit{N} (cm$^{-2}$)            \\     
       & Component 1 ($b$ $\sim$ 4 km s$^{-1}$) & Component 2 ($b$ $\sim$ 5 km s$^{-1}$) \\
\hline
$\cto$ & 13.68 (0.07)                          & 13.57 (0.09)                            \\
$\oon$ & 14.50 (0.05)                          & 14.24 (0.10)                            \\
$\sit$ & 13.21 (0.04)                          & 12.80 (0.16)                            \\
$\fet$ & 12.90 (0.29)                          & 12.60 (0.56)                            \\
\hline
\end{tabular}
\label{tab:j0953}
\end{table}

%
%
\subsection{z$_{abs}$ = 2.5397, 2.68537, 2.74575 towards J1004$+$0018} 
\label{sec_J1004$+$0018}  
\begin{figure*}
\centering
\includegraphics[width=0.8\textwidth, angle=90]{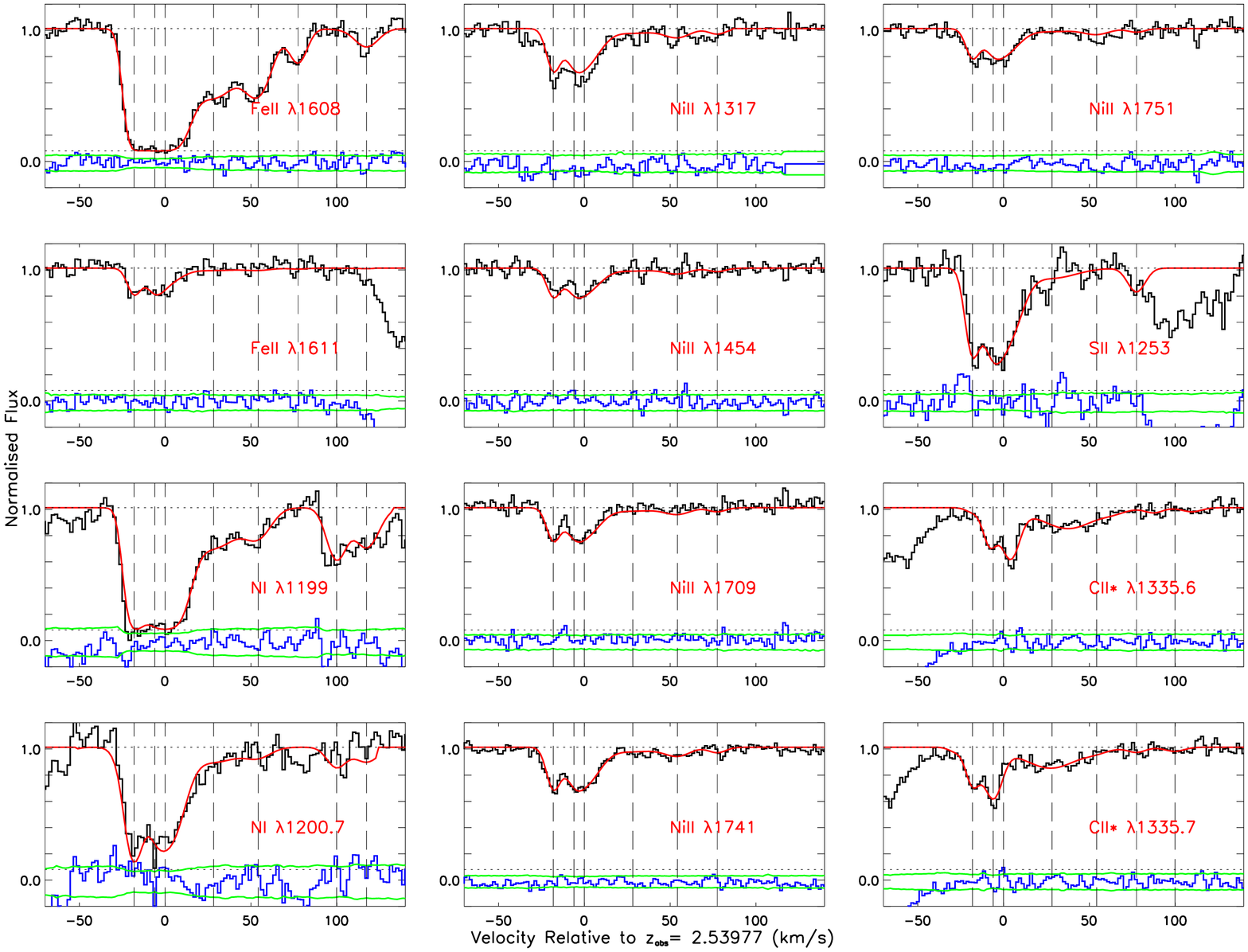}
\caption{Same as in Fig. \ref{fig:ionsj0035} for a selection of metal lines associated with the DLA at z$_{abs}$ = 2.5397 towards J1004$+$0018.}
\label{fig:ionsj1004c}
\end{figure*}

\begin{figure*}
\centering
\includegraphics[width=0.8\textwidth, angle=90]{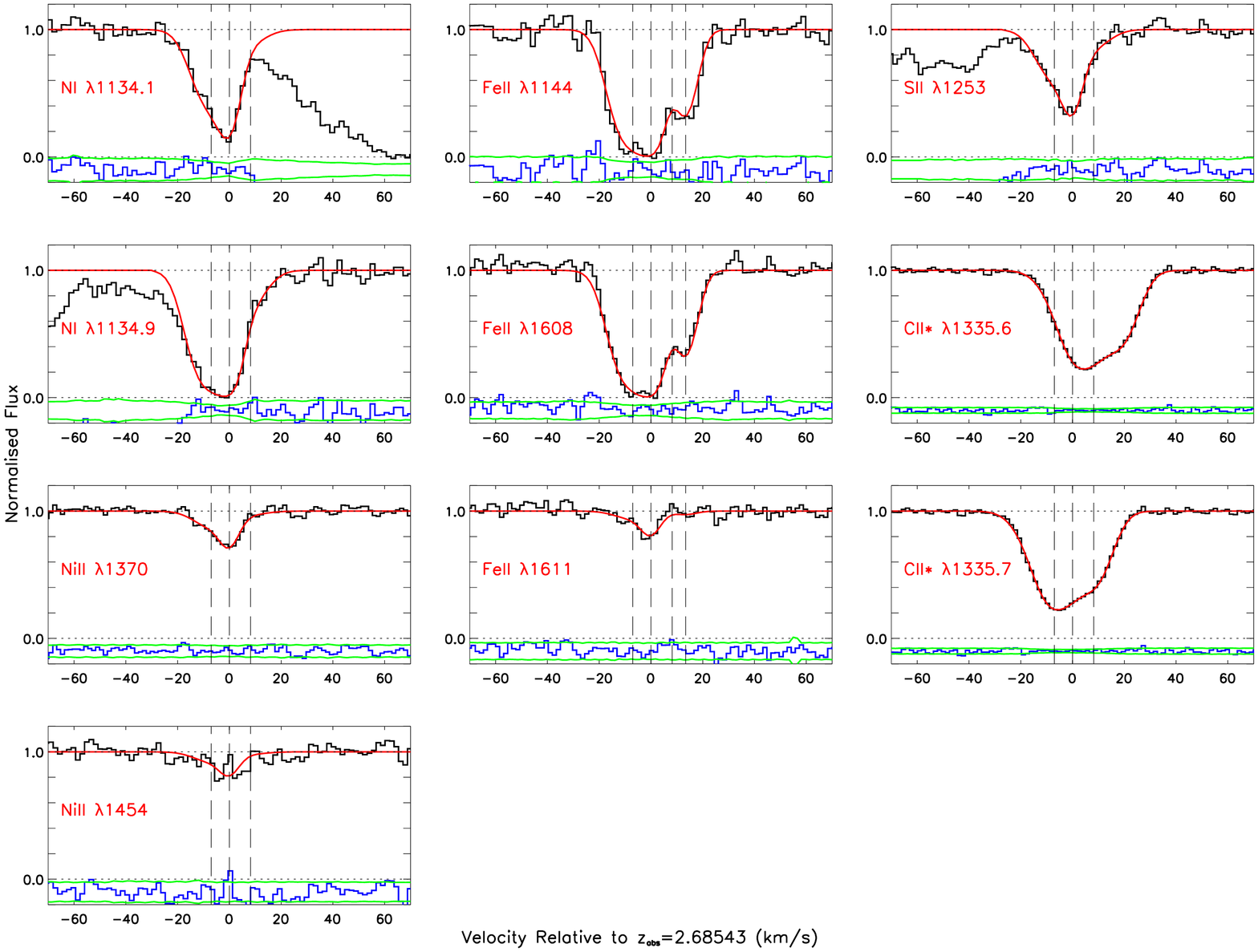}
\caption{Same as in Fig. \ref{fig:ionsj0035} for a selection of metal lines associated with the DLA at z$_{abs}$ = 2.68537 towards J1004$+$0018.}
\label{fig:ionsj1004b}
\end{figure*}

\begin{figure*}
\centering
\includegraphics[width=0.3\textwidth, angle=90]{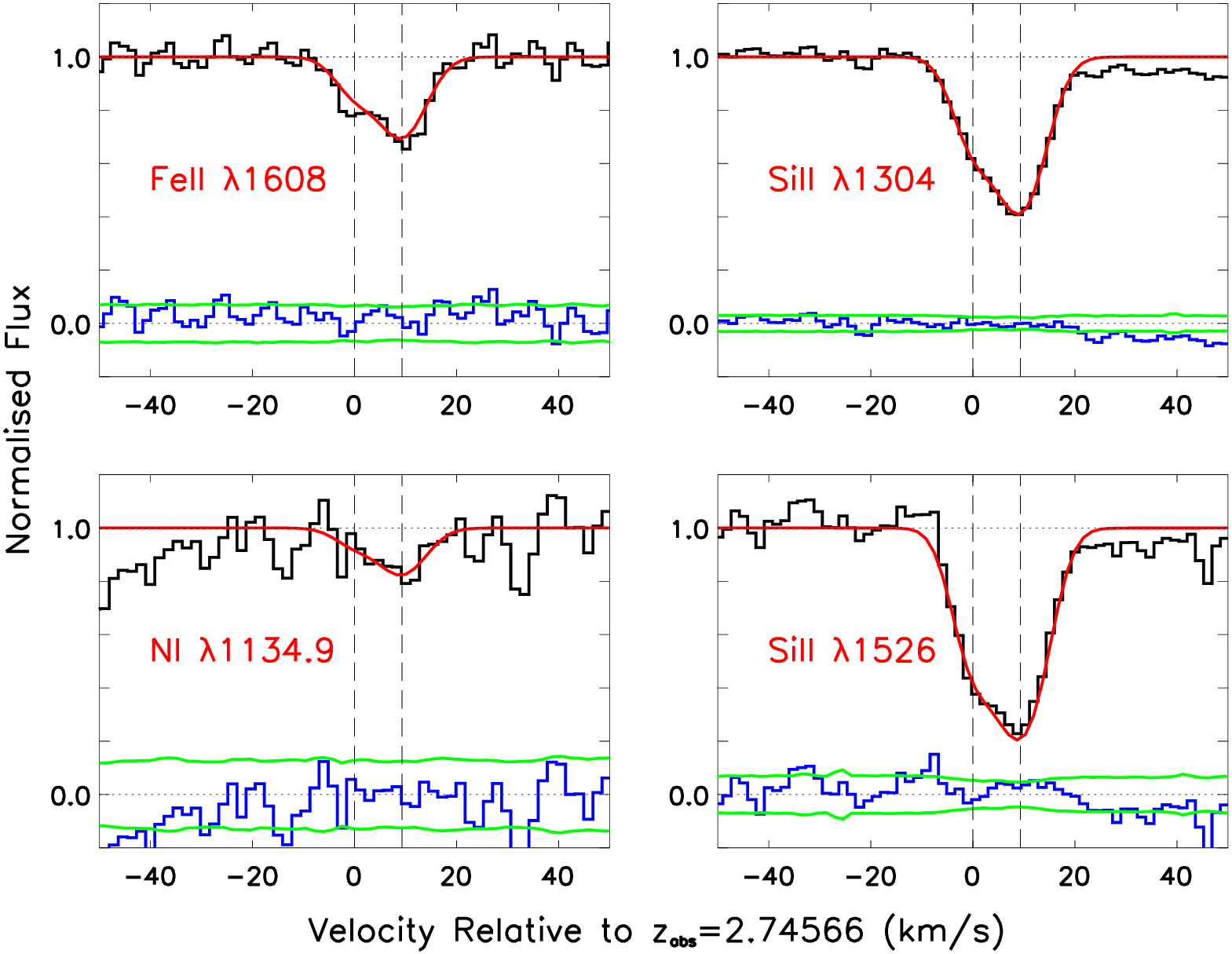}
\caption{Same as in Fig. \ref{fig:ionsj0035} for a selection of metal lines associated with the sub-DLA at z$_{abs}$ = 2.74575 towards J1004$+$0018.}
\label{fig:ionsj1004a}
\end{figure*}

This sightline contains two DLAs and one sub-DLA. They were selected on the basis of the weakness of metal lines in the SDSS
spectrum. It is interesting to note that within a redshift range of $\sim$ 0.2 ($\Delta$v $\sim$ 17000 km s$^{-1}$) of the 
three systems, the [Fe/H] of the systems varies by $\sim$ 0.43 dex. The configuration of three closely spaced systems seen here 
is similar to the few rare cases known in literature \citep{led03,lop03,sri10}. It has been observed that one of the DLAs in
such a configuration happens to be metal-poor, which seems to be the case here as well. \\
\indent We derive log[\textit{N}($\hon$)(cm$^{-2}$)] = 21.30 $\pm$ 0.10 for the DLA at $z_{abs}$ = 2.5397 (see Fig. \ref{fig:dlas}). 
The best-fitting cloud model consists of eight components, not all present in every ion, with $b$ parameters ranging from $\sim$ 4 to $\sim$ 
12 km s$^{-1}$. The metal lines show a more complex and extended component structure than that of any other systems in the present sample 
(see Fig. \ref{fig:ionsj1004c}). Moreover, with [Fe/H] $\sim$ $-$1.7, this is the only DLA in the current study not satisfying our definition 
of metal-poor. The presence of relatively high metallicity as well as larger velocity width of the line profiles is consistent with the 
velocity-metallicity relation in DLAs as observed by \citet{led06}. This being a relatively metal-rich DLA, all the transitions of $\oon$, 
$\cto$ and $\sit$ are heavily saturated. The lower limits to their column densities have been estimated as described in Section \ref{sec_J0234}.
We also detect many transitions of $\nit$ in this system, which are shown in Fig. \ref{fig:ionsj1004c}. From our best fit, we get
log[\textit{N}($\nit$)(cm$^{-2}$)] = 13.91 $\pm$ 0.01. We give the column densities of selected ions in Table \ref{tab:colden} and abundances 
in Table \ref{tab:met}. \citet{jor13} reports log[\textit{N}($\hon$)(cm$^{-2}$)]= 21.10 $\pm$ 0.10 and [Fe/H] = $-$1.23 $\pm$ 0.03 for this system 
without details. While our \textit{N}($\hon$) measurements are consistent within errors, our metallicity estimates differ, due most likely to the medium 
resolution spectrum used by \citet{jor13}. \\
\indent The system at $z_{abs}$ = 2.68537 is the second DLA along this line of sight. It has the highest neutral hydrogen 
column density in the present sample, with log[\textit{N}($\hon$)(cm$^{-2}$)] = 21.39 $\pm$ 0.10 (see Fig. \ref{fig:dlas}). 
Even though this is a low-metallicity system, with [Fe/H] $\sim$ $-$2.13, the strong transitions of $\oon$, $\cto$ and $\sit$ 
covered by our spectrum are saturated. The metal line profiles are best fitted by a three-component cloud model (with $b$ 
parameters between $\sim$ 4 and $\sim$ 8 km s$^{-1}$), though a fourth component is required to fit the Fe profiles (see Fig. 
\ref{fig:ionsj1004b}). $\nit$ is detected in this system and we obtain log[\textit{N}($\nit$)(cm$^{-2}$)] = 13.39 $\pm$ 0.02 from our fit. 
The column densities (or lower limits in case of saturated profiles) of the relevant ions are given in Table \ref{tab:colden}
,while the metallicities are presented in Table \ref{tab:met}. For this system also, the \textit{N}($\hon$) measurement 
reported by \citet{jor13} (log[\textit{N}($\hon$)(cm$^{-2}$)] = 21.25 $\pm$ 0.10), using medium resolution spectrum, match with ours within 
errors, while their abundance measurement ([Fe/H] = $-$1.73 $\pm$ 0.06) differs from that obtained by us. The systems at 
z$_{abs}$ = 2.5397 and 2.68537 are the two in our sample that clearly show $\cto$* absorption. We discuss the implications of the 
$\cto$* detections in Section \ref{sec_ctos}. \\
\indent The sub-DLA at $z_{abs}$ = 2.74575 along this sightline, with [Fe/H] $\sim$ $-$2, is also a low-metallicity system. 
We deduce log[\textit{N}($\hon$)(cm$^{-2}$)] = 19.84 $\pm$ 0.10 (see Fig. \ref{fig:dlas}), and fit a two component cloud model 
(with $b$ parameters of 4.11 $\pm$ 0.48 and 5.29 $\pm$ 0.34 km s$^{-1}$ separated by $\sim$ 9.3 km s$^{-1}$) to the metal lines 
(see Fig. \ref{fig:ionsj1004a}). The gas in sub-DLAs may be partially ionized, however ionizations corrections may not be very 
important at lower metallicities \citep{per07}. Here, we have not applied any ionization corrections while deriving the abundances. 
In this system also, the $\oon$ $\lambda$1302 and $\cto$ $\lambda$1334 lines are saturated and blended, and hence we provide only 
lower limits to their column densities. We also compute the 3$\sigma$ upper limit to the $\sto$ column density using the strongest 
undetected transition ($\lambda$1259), as per the method described in Section \ref{sec_J0035}. Table \ref{tab:colden} lists the 
column densities of the relevant ions and Table \ref{tab:met} the respective abundances. 
\begin{table*} 
\caption{Total column densities of selected ions in the DLAs}
\centering
\begin{tabular}{ccccccccc}
\hline
QSO          & z$_{abs}$ & log~\textit{N}($\hon$) & log~\textit{N}($\cto$) & log~\textit{N}($\non$) & log~\textit{N}($\oon$) & log~\textit{N}($\sit$) & log~\textit{N}($\sto$) & log~\textit{N}($\fet$) \\
             &           & (cm$^{-2}$)            & (cm$^{-2}$)            & (cm$^{-2}$)            & (cm$^{-2}$)            & (cm$^{-2}$)            & (cm$^{-2}$)            & (cm$^{-2}$)            \\
\hline
J0035$-$0918 & 2.34006   & 20.55 (0.10)           & 14.45 (0.19)           & 13.48 (0.03)           & 14.55 (0.14)           & 13.37 (0.06)           & $\le$ 13.13            & 13.07 (0.04)           \\      
J0234$-$0751 & 2.31815   & 20.90 (0.10)           & $\ge$ 13.80            & 14.23 (0.03)           & $\ge$ 14.25            & 14.32 (0.09)           & 14.18 (0.03)           & 14.18 (0.03)           \\
J0953$-$0504 & 4.20287   & 20.55 (0.10)           & 13.93 (0.02)           & $\le$ 13.54            & 14.69 (0.02)           & 13.35 (0.02)           & $\le$ 13.89            & 13.07 (0.19)           \\
J1004$+$0018 & 2.53970   & 21.30 (0.10)           & $\ge$ 14.54            & 14.73 (0.04)           & $\ge$ 14.91            & $\ge$ 14.33            & 15.09 (0.01)           & 15.13 (0.02)           \\
J1004$+$0018 & 2.68537   & 21.39 (0.10)           & $\ge$ 14.02            & 14.86 (0.02)           & $\ge$ 14.54            & $\ge$ 13.87            & 14.70 (0.02)           & 14.71 (0.04)           \\
J1004$+$0018 & 2.74575   & 19.84 (0.10)           & $\ge$ 13.74            & 13.35 (0.07)           & $\ge$ 14.25            & 13.67 (0.01)           & $\le$ 13.40            & 13.31 (0.02)           \\
\hline
\end{tabular}
\label{tab:colden}
\end{table*}
%
%
\section{Elemental abundances in metal-poor DLAs} 
\label{sec_met}
\begin{figure*}
\includegraphics[width=0.6\textwidth, angle=90]{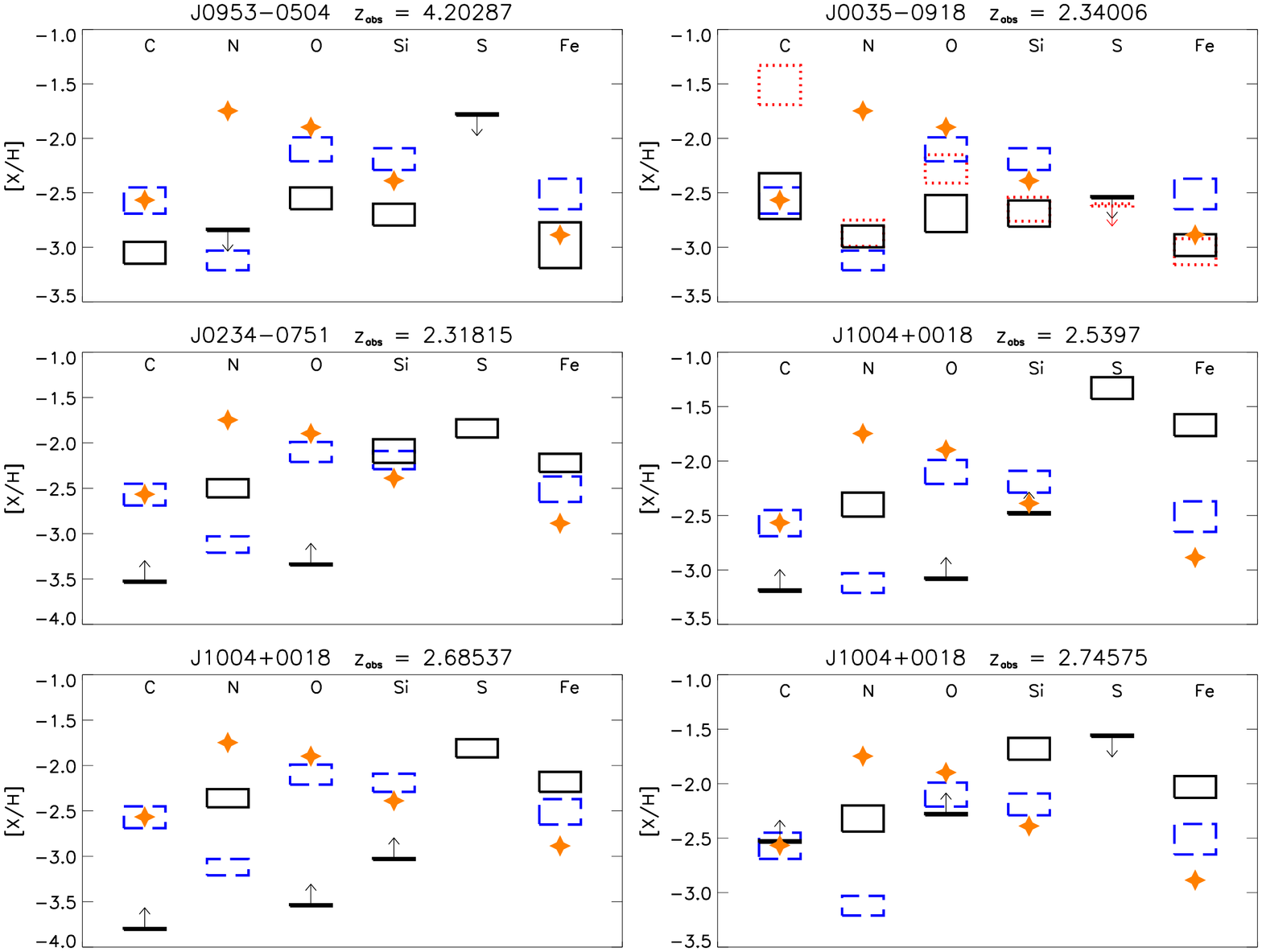}
\caption{Elemental abundances in the DLAs. The height of each box represents the uncertainty in each elemental abundance.
Upper and lower limits are indicated by bar and arrow. The black boxes show the abundances measured for our sample; 
the dashed blue boxes represent the abundance pattern of a typical VMP DLA \citep{cookb11}; the dotted red boxes in 
the upper right panel show the abundances of the CEMP DLA at z$_{abs}$ = 2.34006 towards J0035$-$0918 as reported by 
\citet{cooka11}; the stars show the typical abundance pattern of metal-poor halo stars obtained from \citet{cay04} 
and \citet{beer05}. QSO names and absorption redshifts are provided in each panel.}
\label{fig:comp}
\end{figure*}

\indent Typically metal-poor DLAs seem to roughly follow a similar abundance pattern. Deviations in the abundance pattern are 
seen for CEMP DLAs, which are usually understood as enrichment by core-collapse supernovae of massive primordial stars \citep{kob11}. 
In our sample, except for the CEMP DLA towards J0035$-$0918, all the systems with [Fe/H] $\le$ $-$2 seem to follow the typical relative abundance pattern 
of elements in a metal-poor DLA, indicating that the elemental abundances in such DLAs arise from similar population of stars. 
In Table \ref{tab:met}, we list the abundances of the relevant elements in our sample, and in Fig. \ref{fig:comp}, we present a 
graphical comparison of the abundances measured in our systems (solid boxes) with those of a typical VMP DLA as defined by \citet{cookb11} (dashed boxes).
We also compare the measured abundances with that of typical metal-poor stars, which we obtain by taking the average of the 
abundances in metal-poor stars in the halo of the Galaxy (with [Fe/H] ranging between $-$2.0 and $-$3.5), as given by \citet{cay04} 
and \citet{beer05}. We find that the abundance pattern of metal-poor stars (shown as stars) shows greater enhancement of C, N and O with 
respect to Fe than that seen till now in metal-poor DLAs, in particular, the abundance of N is more than a magnitude higher 
in such stars than seen in metal-poor DLAs (see Fig. \ref{fig:comp}). The top right panel of Fig. \ref{fig:comp} also 
shows the comparison of the elemental abundances in the DLA at z$_{abs}$ = 2.34006 towards J0035$-$0918, obtained by us with 
that by \citet{cooka11} (dotted boxes). Our results are found to be consistent with theirs, except for that of carbon and oxygen. As discussed 
in Section \ref{sec_J0035}, this is because while \citet{cooka11} consider turbulence as the broadening mechanism behind the 
line widths (which is equal to assuming T = 0 K), we adopt the results obtained from thermal broadening (T $\sim$ 8000 K). \\
\indent In the following sections, we discuss the abundances of C, N, O in metal-poor DLAs and the overall trend of their abundance
ratios as a function of metallicity, by combining our sample with that present in literature. These elements play the most important 
role in the nucleosynthesis of stars, and hence studying their relative abundance patterns at low-metallicity is crucial to understand 
the composition and yields of the first few generations of stars. 
\begin{table*} 
\caption{Abundance measurements of selected elements in the DLAs}
\centering
\begin{tabular}{ccccccccc}
\hline
QSO          & z$_{abs}$ & log~\textit{N}($\hon$) & [C/H]          & [N/H]          & [O/H]          & [Si/H]         & [S/H]          & [Fe/H]         \\
             &           & (cm$^{-2}$)            &                &                &                &                &                &                \\
\hline
J0035$-$0918 & 2.34006   & 20.55 (0.10)           & $-$2.53 (0.21) & $-$2.90 (0.10) & $-$2.69 (0.17) & $-$2.69 (0.12) & $\le$ $-$2.54  & $-$2.98 (0.10) \\       
J0234$-$0751 & 2.31815   & 20.90 (0.10)           & $\ge$ $-$3.53  & $-$2.50 (0.10) & $\ge$ $-$3.34  & $-$2.09 (0.13) & $-$1.84 (0.10) & $-$2.22 (0.10) \\
J0953$-$0504 & 4.20287   & 20.55 (0.10)           & $-$3.05 (0.10) & $\le$ $-$2.84  & $-$2.55 (0.10) & $-$2.70 (0.10) & $\le$ $-$1.78  & $-$2.98 (0.21) \\
J1004$+$0018 & 2.53970   & 21.30 (0.10)           & $\ge$ $-$3.19  & $-$2.40 (0.11) & $\ge$ $-$3.08  & $\ge$ $-$2.48  & $-$1.33 (0.10) & $-$1.67 (0.10) \\
J1004$+$0018 & 2.68537   & 21.39 (0.10)           & $\ge$ $-$3.80  & $-$2.36 (0.10) & $\ge$ $-$3.54  & $\ge$ $-$3.03  & $-$1.81 (0.10) & $-$2.18 (0.11) \\
J1004$+$0018 & 2.74575   & 19.84 (0.10)           & $\ge$ $-$2.53  & $-$2.32 (0.12) & $\ge$ $-$2.28  & $-$1.68 (0.10) & $\le$ $-$1.56  & $-$2.03 (0.10) \\
\hline
\end{tabular}
\label{tab:met}
\end{table*}
%
%
\subsection{The O/Fe Ratio}
\label{sec_ofe}
Oxygen is mainly produced by massive stars that undergo Type II supernova explosions (SNe II), whereas, iron is mostly 
(i.e. 2/3$^{rd}$) produced by Type Ia supernova explosions (SNe Ia) of low and intermediate mass stars. Since low and 
intermediate mass stars take longer time to evolve than massive stars, SNe Ia occur $\sim$ 1 Gyr later than SNe II. 
Therefore, initially at low metallicity, oxygen is expected to be enhanced relative to iron. Later, when the delayed 
contribution to iron by SNe Ia starts, the O/Fe ratio is expected to decrease. Thus, the relative abundance of O and Fe 
is a good indicator of the relative contribution of SNe Ia and SNe II yields towards chemical enrichment, as well as the 
time delay between them. Additionally, at low metallicity the O/Fe ratio can serve as a measure of the relative production of
$\alpha$ to iron-peak elements by the early generations of high mass stars.\\
\indent The measurement of oxygen abundance in DLAs is usually an issue, since the stronger $\oon$ lines are mostly saturated 
and the weaker ones are often blended within the Ly$\alpha$ forest. As can be seen from Table \ref{tab:met}, we have direct 
measurement of the oxygen abundance for only two DLAs. For the DLA at z$_{abs}$ = 2.34006 towards J0035-0918, we get [O/Fe] = 
0.29 $\pm$ 0.15, while for the other DLA at z$_{abs}$ = 4.20287 towards J0953$-$0504, we find [O/Fe] = 0.43 $\pm$ 0.20. These 
are among the four measurements of oxygen in DLAs with [Fe/H] $\le$ $-$2.9. Our values do not seem to imply an upward 
trend of [O/Fe] at [Fe/H] $<$ $-$3.0, as hinted at by \citet{cookb11}, however more data are required to reach any definite 
conclusions. For remaining systems we use S (or Si) as proxy to O abundance measurements using solar abundances (abundances of S 
and Si follow that of O, all being $\alpha$-capture elements). The average [$\langle$O/Fe$\rangle$] for the five systems in our 
sample with [Fe/H] $\le$ $-$2.0 is 0.36 $\pm$ 0.26. This is consistent with the average [$\langle$O/Fe$\rangle$] = 0.35 $\pm$ 0.09 
found in DLAs with $-$3 $\le$ [Fe/H] $\le$ $-$2, as well as the [$\langle$O/Fe$\rangle$] $\sim$ 0.4 seen in metal-poor stars with 
$-$3.5 $\le$ [Fe/H] $\le$ $-$1 \citep{cookb11}. The [O/Fe] ratios for our sample are shown in the top right panel of Fig. \ref{fig:ratio}, 
along with those for VMP DLAs in literature, as well as for metal-poor halo stars as given in \citet{cookb11}, for comparison. 
It has been shown that DLAs typically exhibit minimal dust depletion when [Fe/H] $\lesssim$ $-$2.0 \citep{led03,vla04, akr05}. 
\citet{cay04} find constant [Si/Fe] $\sim$ 0.37 from the measurements of metal-poor stars in the halo of the Galaxy, indicating 
that at lower metallicities the intrinsic nucleosynthetic Si/Fe ratio is almost independent of the metallicity. 
For our present sample, we find that [$\langle$Si/Fe$\rangle$] $\sim$ 0.34, consistent with that found by \citet{cay04}.
%
%
\subsection{The N/O Ratio}
\label{sec_no}
Nitrogen is mainly produced through the CNO cycle in hydrogen burning layers of stars. It is believed to be of both
primary and secondary origin, depending on whether the seed C and O are produced by the star itself during
helium burning (primary), or whether they are leftovers from earlier generations of stars and hence already present in 
the interstellar medium (ISM) from which the star formed (secondary). Primary N is thought to be generated by intermediate 
mass stars on the asymptotic giant branch (AGB). Secondary N is produced by all stars as C and O are present in their H burning layers.
In $\hto$ regions of nearby galaxies, it has been observed that the N/O ratio rises steeply with increasing oxygen abundance 
for [O/H] $\gtrsim$ $-$0.4; this is the secondary regime. At lower metallicity, for [O/H] $\lesssim$ $-$0.7, the N/O ratio 
remains constant; this is the primary regime where N abundance tracks that of O.\\
\indent The measurement of N/O ratio in DLAs is again complicated by the fact that the $\oon$ absorption lines are either
saturated or blended, while the $\non$ lines may be blended in the Ly$\alpha$ forest. Here, we adopt the standard practice 
of replacing O abundance by that of another $\alpha$ element S (or Si), whenever measurement of O is not available. In upper 
right panel of Fig. \ref{fig:ratio}, we plot the [N/O] ratio versus O abundance for our sample, as well as values reported by 
\citet{petj08} and \citet{cookb11}. We also show the location of the local primary plateau ([N/O] between $-$0.57 and $-$0.74) 
and secondary production region (extrapolated from local measurements), as given by \citet{petj08}. Most of the DLA values lie 
within these two regions, which may imply that the DLAs are in the transition period following a burst of star formation, when 
the ISM has been enriched by O released from SNe II, but lower mass stars have yet to release their primary N. To this group belongs 
the system at z$_{abs}$ = 2.5397 towards J1004$+$0018 studied here. In one of the two cases (z$_{abs}$ = 4.20287 towards J0953$-$0504) 
where we have O measurement, we could only get upper limit on \textit{N}($\non$). This suggests [N/O] $\le$ $-$0.29, which is above 
the primary plateau. So nothing definitive can be said about the relative enrichment of N and O in this system. For the system at 
z$_{abs}$ = 2.34006 towards J0035$-$0918, we get [N/O] = $-$0.21 $\pm$ 0.14. This lies above the local primary plateau, i.e., the 
amount of nitrogen relative to oxygen in this system is higher than what is seen in typical DLAs. Note that this is the CEMP
VMP DLA whose abundance pattern shows deviations from that of a typical VMP DLA (see Fig. \ref{fig:comp}). The [N/O] ratio for three of 
the remaining systems (where we have used S or Si as proxy for O) fall along the primary plateau. The high N/O ratios in these systems 
indicate that the release of primary N by intermediate mass AGB stars into the ISM has caught up with that of O by massive stars.
%
%
\subsection{The C/O Ratio}
\label{sec_co}
\begin{figure*}
\includegraphics[width=0.3\textwidth, angle=90]{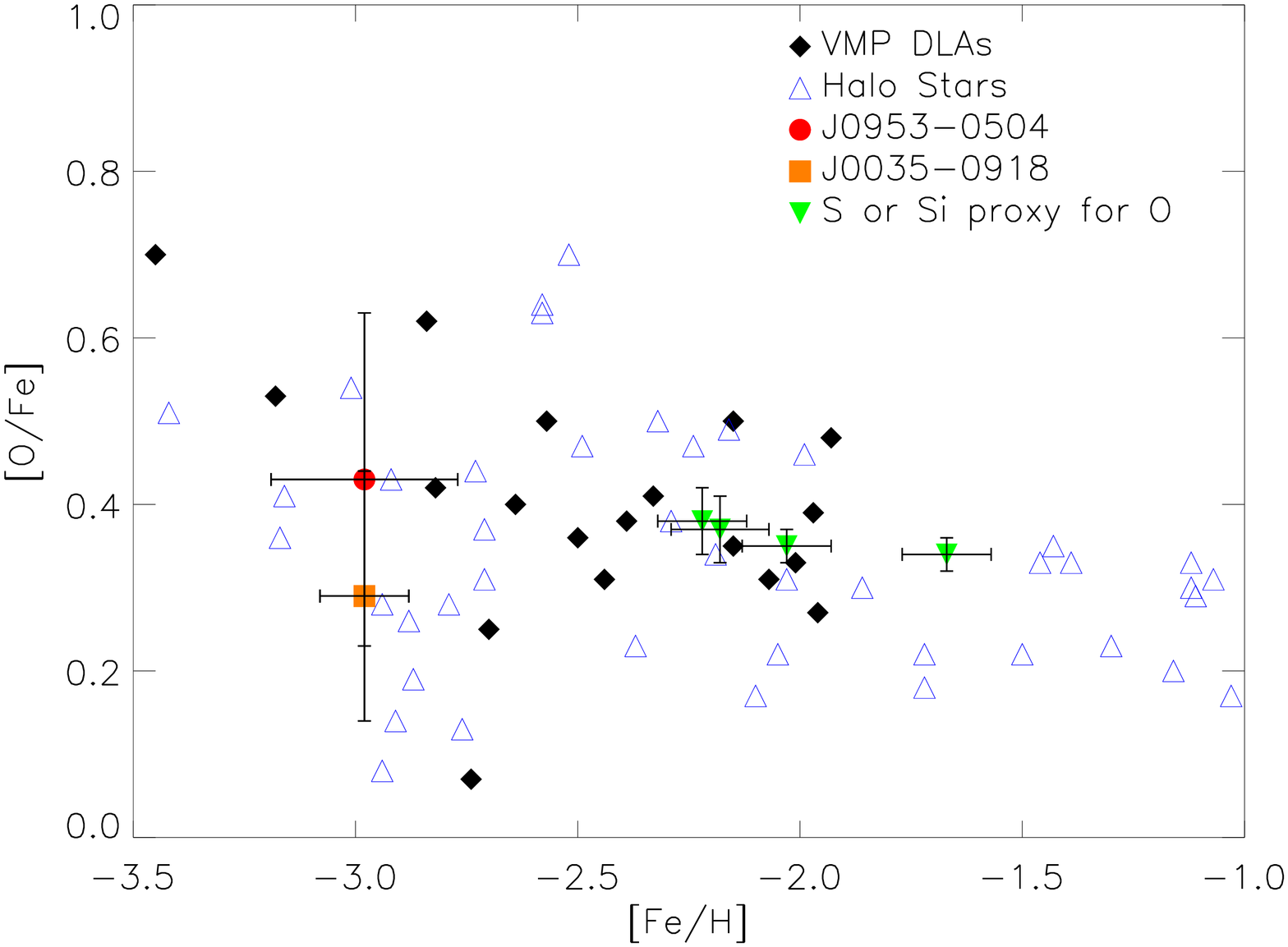}
\includegraphics[width=0.3\textwidth, angle=90]{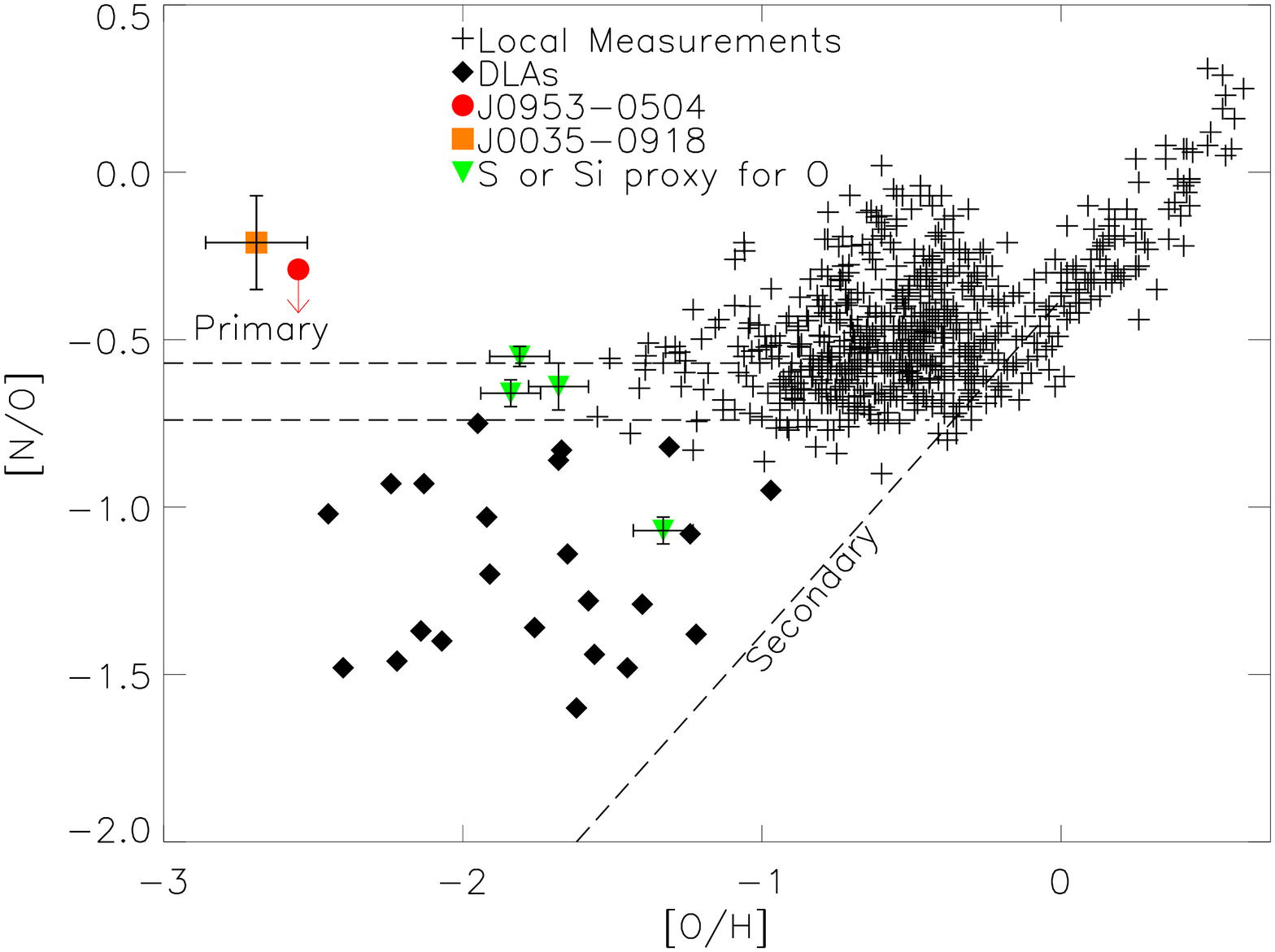}
\centering \includegraphics[width=0.3\textwidth, angle=90]{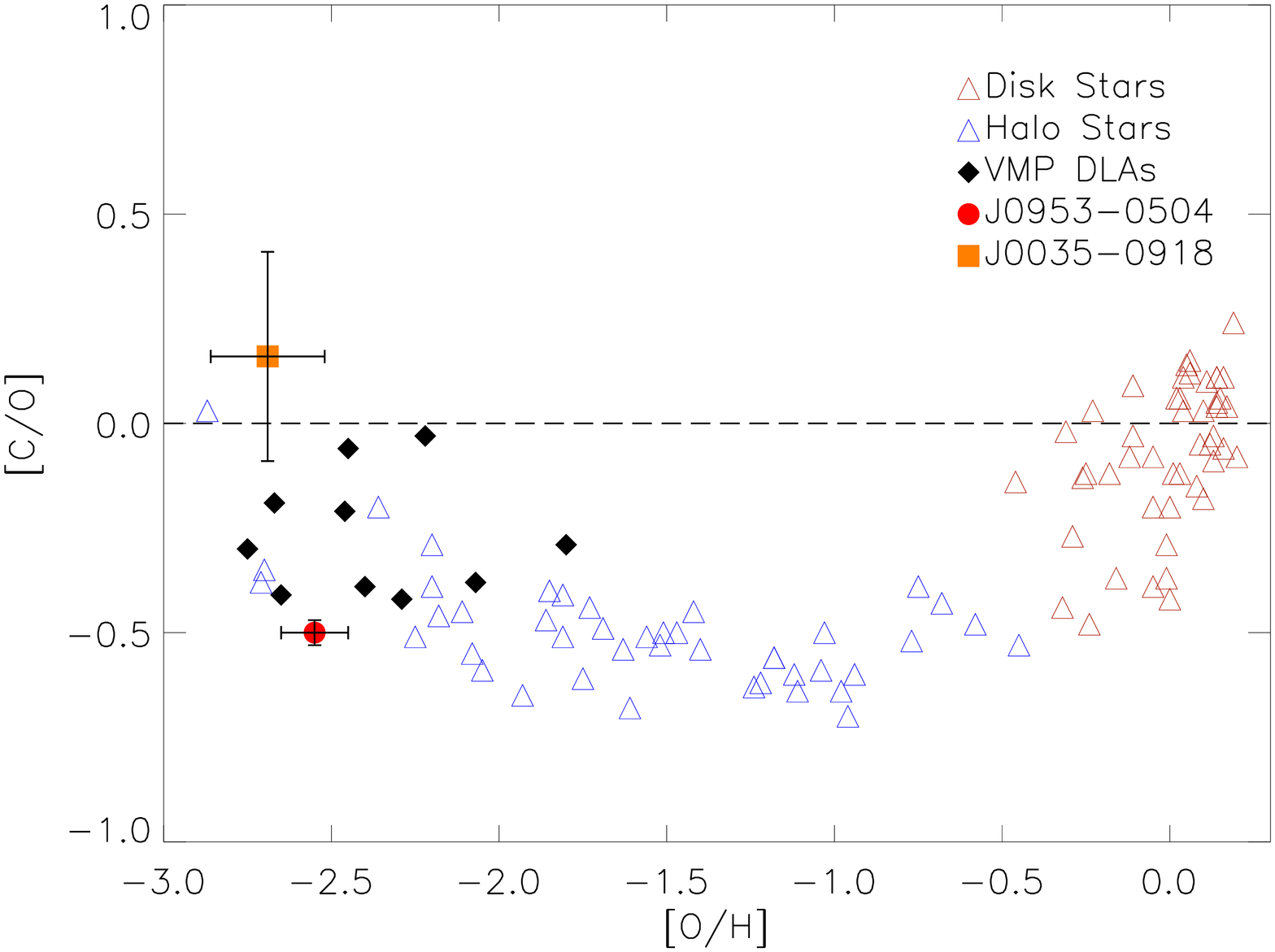}
\caption{\textit{Top Right:} The [O/Fe] ratio Vs. [Fe/H] in our DLAs compared with that of VMP DLAs and metal poor stars as compiled by \citet{cookb11}.
\textit{Top Left:} The [N/O] ratio Vs. [O/H] in our DLAs compared with that of DLAs and local measurements as given by \citet{petj08} and 
\citet{cookb11} and sources cited by them. 
\textit{Bottom:} The [C/O] ratio Vs. [O/H] in our DLAs compared with that of VMP DLAs and metal poor stars as compiled by \citet{cookb11}.}
\label{fig:ratio}
\end{figure*}

The evolution of the [C/O] ratio with O abundance has been studied by \citet{akr04} for halo stars, who found that [C/O] rises
from $\sim$ $-$0.5 to solar when [O/H] $\gtrsim$ $-$1. This is interpreted as the the additional contribution to C, from massive 
stars whose mass loss rates increase with metallicity, and less importantly, from low and intermediate mass stars that take 
longer to evolve than massive stars. More interestingly, \citet{akr04} found an increasing trend of [C/O] with decreasing metallicity
when [O/H] $\lesssim$ $-$1, suggesting that [C/O] may reach near-solar values when [O/H] $\lesssim$ $-$3. This trend of rising [C/O]
with decreasing O abundance has also been observed in low-metallicity DLAs by \citet{petn08}, \citet{penp10} and \citet{cookb11}, 
indicating the enhancement of C over O in this regime may have a universal origin. Models of Population II nucleosynthesis predict 
that below [O/H] $\sim$ $-$1, [C/O] decreases (or perhaps flattens) with decreasing metallicity (due to time lag in C production 
relative to O), in contrast to the observed trend.\\
\indent The upward trend in [C/O] with decreasing metallicity can be explained either as signatures of high carbon production by 
the first generation Population III stars, or as increased carbon yield from rapidly rotating low-metallicity Population II stars. 
Fig. \ref{fig:ratio} (bottom panel) shows the [C/O] ratio versus [O/H] for metal-poor stars and DLAs as given in \citet{cookb11}. 
Oxygen and carbon both have very strong transitions which are usually saturated in DLAs. In the present study we have reliable 
abundance measurement of O and C in only two systems, which we show in Fig. \ref{fig:ratio}. For the DLA at z$_{abs}$ = 4.20287 
towards J0953$-$0504, the abundance of O is low ([O/H] $\lesssim$ $-$2.5), however C is not enhanced with respect to O ([C/O] = $-$0.5 
$\pm$ 0.03). At least in this system, abundances of C and O are not following what is expected in the case of chemical evolution 
dominated by Population III stars. No DLA has been observed with super-solar [C/O], apart from the system (towards J0035$-$0918) 
reported by \citet{cooka11} and also analysed here. However, as discussed in section \ref{sec_J0035}, the [C/O] ratio obtained by us 
is 0.16 $\pm$ 0.25, about four times less than 0.77 $\pm$ 0.17 obtained by \citet{cooka11}. While the amount of enhancement 
that we find is smaller compared to that reported by \citet{cooka11}, the enhancement of C still indicates that the gas in this 
system may have been enriched by yields of Population III stars.
%
%
\section{$\cto$* in metal-poor DLAs} 
\label{sec_ctos}
\begin{figure}
\centering
\includegraphics[totalheight=0.3\textheight, angle=90]{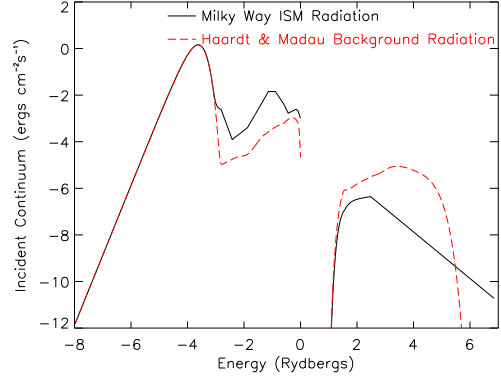}
\caption{The incident continuum spectra used in our photoionization models.}
\label{fig:cont}
\end{figure}

The $\cto$* $\lambda$1335 line arises due to transition from $^{2}$P$_{3/2}$ state to higher lying $^{2}$D$_{3/2}$ and
$^{2}$D$_{5/2}$ states in C$^{+}$. The C$^{+}$ fine-structure transition can be excited due to collisions with electrons, 
protons, or H$_{2}$ molecules, depending on the physical conditions prevailing in the gas. The [$\cto$] $\lambda$158 $\mu$m 
line results from radiative deexcitation between the $^{2}$P$_{3/2}$ and $^{2}$P$_{1/2}$ fine-structure states in the ground 
2s$^{2}$2p term of C$^{+}$. In the Galactic ISM at T $<$ 300 K, radiative cooling is dominated by emission of the fine-structure 
line [$\cto$] $\lambda$158 $\mu$m \citep{wolfr95}. The [$\cto$] $\lambda$158 $\mu$m cooling rate per H atom is given by,
\begin{equation}
l_{c}  = \frac{ \textit{N}(\cto^{*}) h\nu_{ul} A_{ul} } { \textit{N}(\hon) }  ~ergs~s^{-1}~H^{-1} 
\label{eqn:cool}
\end{equation}
where, \textit{N}($\cto$*) is the column density of C$^{+}$ ions in the $^{2}$P$_{3/2}$ state, h$\nu_{ul}$ (92 K) and A$_{ul}$ 
(2.4 $\times$ 10$^{-6}$ s$^{-1}$) are the energy and coefficient for spontaneous photon decay of the $^{2}$P$_{3/2}$ 
$\rightarrow$ $^{2}$P$_{1/2}$ transition respectively. \\
\indent We observe $\cto$* $\lambda$1335 absorption in two of our systems, along the sightline J1004$+$0018: log[\textit{N}($\cto$*)(cm$^{-2}$)] = 
13.63 $\pm$ 0.02 in the DLA at z$_{abs}$ = 2.5397 (hereafter DLA1); and log[\textit{N}($\cto$*)(cm$^{-2}$)] = 13.94 $\pm$ 0.03 in the DLA at z$_{abs}$ = 
2.68537 (hereafter DLA2). For DLA1, we compute log(\textit{l$_{c}$}) to be $-$27.19 $\pm$ 0.10 ergs s$^{-1}$ H$^{-1}$, while it is 
$-$26.97 $\pm$ 0.10 for DLA2 (see Table \ref{tab:ctos1}). These values fall in the trough of the bimodal distribution of \textit{l$_{c}$} 
in DLAs as reported by \citet{wolf08}. They propose two separate populations of  DLAs : ``low cool'' DLAs with \textit{l$_{c}$} $\le$ 
\textit{l$_{c}^{crit}$} and ``high cool'' DLAs with \textit{l$_{c}$} $>$ \textit{l$_{c}^{crit}$} (\textit{l$_{c}^{crit}$} $\simeq$ 10$^{-27}$ 
ergs s$^{-1}$ H$^{-1}$). Both these populations need local radiation in excess of the UV background. While ``low cool'' DLAs are consistent 
with gas embedded in a region with in-situ star formation, ``high cool'' DLA gas may be related to the outskirts of star-forming galaxies. \\
\indent It is well known, for the range of densities seen in the neutral ISM, dust photo-electric heating, cosmic ray (CR) heating, X-ray heating 
and heating by $\con$ ionization are the dominant heating processes \citep{wolfr95}. \citet{wolf03a} have shown that for the \textit{N}($\hon$) 
typically seen in DLAs, the X-ray heating is sub-dominant compared to the grain photo-electric heating and the CR heating. To understand the 
importance of these physical processes in the present low-metallicity systems we construct photoionization models using CLOUDY. \\
\indent We model the DLAs as plane-parallel slabs of gas of constant density exposed to different radiation fields: metagalactic UV background 
radiation given by \citet{hm01} and the ISM radiation field in-built in CLOUDY (i.e spectral energy distribution given in ``Table ISM'').
We consider a range of local ionizing radiation by scaling the ISM radiation with a scale factor (with the strength of unscaled ISM radiation 
field G$_{0}$ corresponding to a flux of 1.6 $\times$ 10$^{-3}$ erg cm$^{-2}$s$^{-1}$).
In Fig. \ref{fig:cont}, we show the two ionizing spectra used in our models. Note when we scale the ISM field, the UV photons as well as X-ray 
background scales by the same factor. Inclusion of cosmic rays is very important as they are known to influence the ionization and chemical 
state of high density ISM. We also include the CR background (as seen in the Galaxy) that corresponds to a H$^0$ ionization rate of 
5.0 $\times$10 $^{-17}$ s$^{-1}$. For simplicity, we scale the CR background by the same amount as UV background in our models. In the case 
of models with Haardt \& Madau background, we include the CR background that is 1/10$^{th}$ of the Galactic CR background.
Note that, we use the extinguish command of CLOUDY to take into account extinction due to photo-electric absorption by a cold neutral slab with 
\textit{N}($\hon$) = 10$^{21}$ cm$^{-2}$. This is the standard procedure used to model ISM sightlines \citep{sha06}. The cosmic microwave 
background at any particular redshift is also included. \\
\indent The dust-to-gas ratio (relative to the Solar System), $\kappa$, is calculated using abundances of a dust-depleted (e.g. Fe) and 
non-depleted (e.g. Zn, Si, S) element as,
\begin{equation}
\kappa = 10^{[X/H]} [ 1 - 10^{[Fe/X]}].
\label{eqn:dust}
\end{equation}
In the present study, we take X $\equiv$ S. Note, the $\kappa$ we measure should be considered an upper limit on the dust content in the case of 
metal-poor DLAs, as relative abundance of Fe with respect to S is most likely due to nucleosynthesis origin (see Section \ref{sec_ofe}). 

For most of the results of photoionization models presented below, we consider radiation field to be impinging on one side of the cloud and
our line of sight to be perpendicular to the slab (i.e., \textit{N}($\hon)^{\perp}$ = \textit{N}($\hon$) observed). However, in the end we discuss the 
results of relaxing these conditions. We also discuss what happens if we assume the absorbing gas to be a parallel slab with constant pressure.
\subsection{Photoionization model for two systems with $\cto$* detections:}
\begin{figure*}
\centering
\includegraphics[totalheight=0.5\textheight, angle=90]{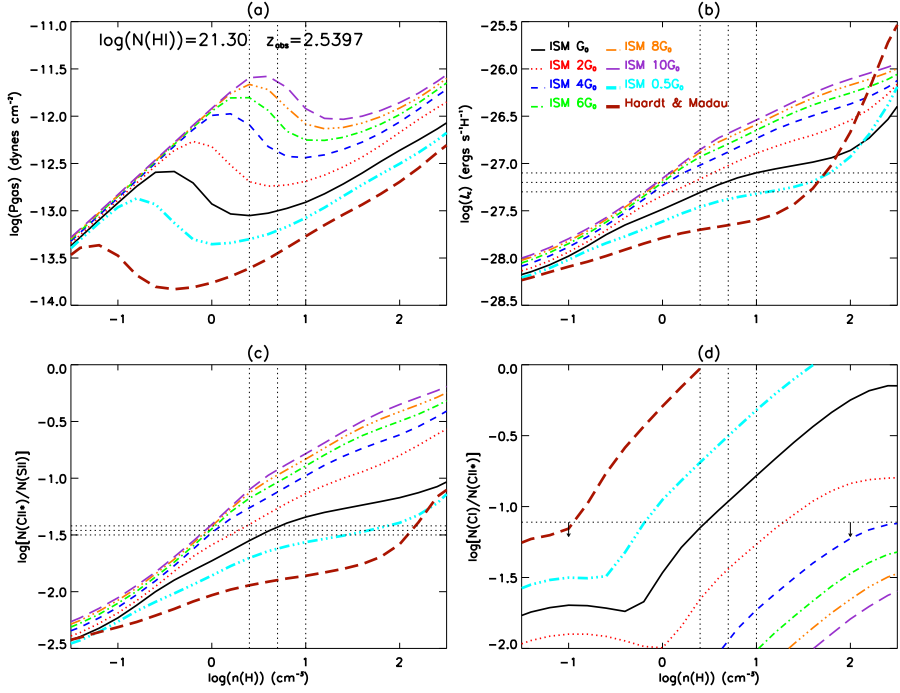}
\caption{(a) Phase diagrams, (b) $\cto$* cooling rates, (c) \textit{N}($\cto$*)/\textit{N}($\sto$) and (d) 
\textit{N}($\con$)/\textit{N}($\cto$*) from CLOUDY for the DLA at z$_{abs}$ = 2.5397 towards J1004$+$0018, for 
various ionizing backgrounds. The horizontal dashed lines indicate the observed values (upper limit in (d)) along 
with the errors. The vertical dashed lines indicate the expected n(H) along with errors for a CNM solution obtained 
from the observed \textit{l$_{c}$} and \textit{N}($\cto$*)/\textit{N}($\sto$).}
\label{fig:dla1}
\end{figure*}

\begin{figure*}
\centering
\includegraphics[totalheight=0.5\textheight, angle=90]{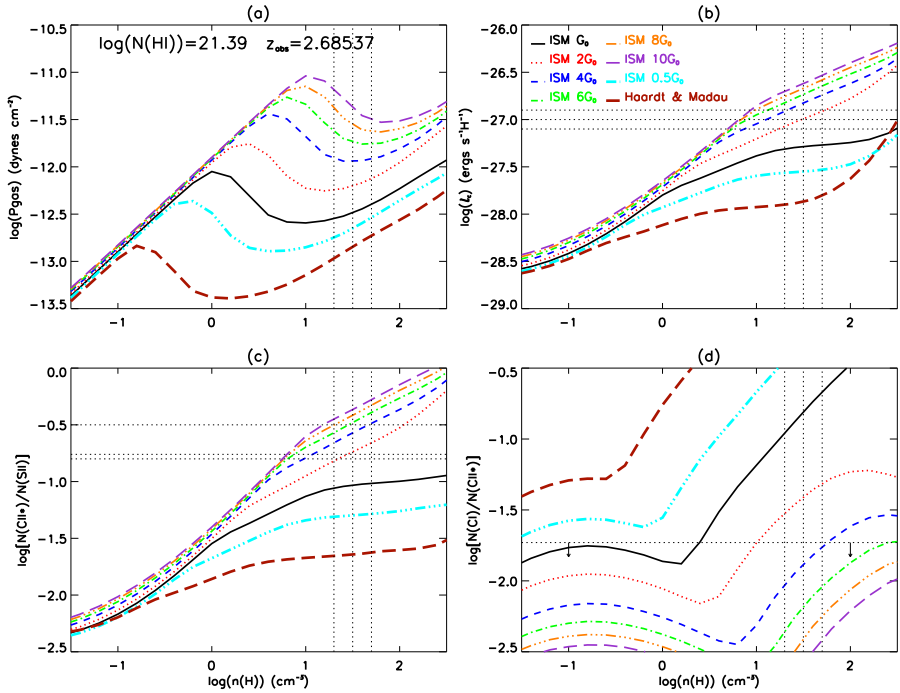}
\caption{The same as in Fig.\ref{fig:dla1} for the DLA at z$_{abs}$ = 2.68537 towards J1004$+$0018}
\label{fig:dla2}
\end{figure*}

\begin{figure}
\includegraphics[width=0.3\textwidth, angle=90]{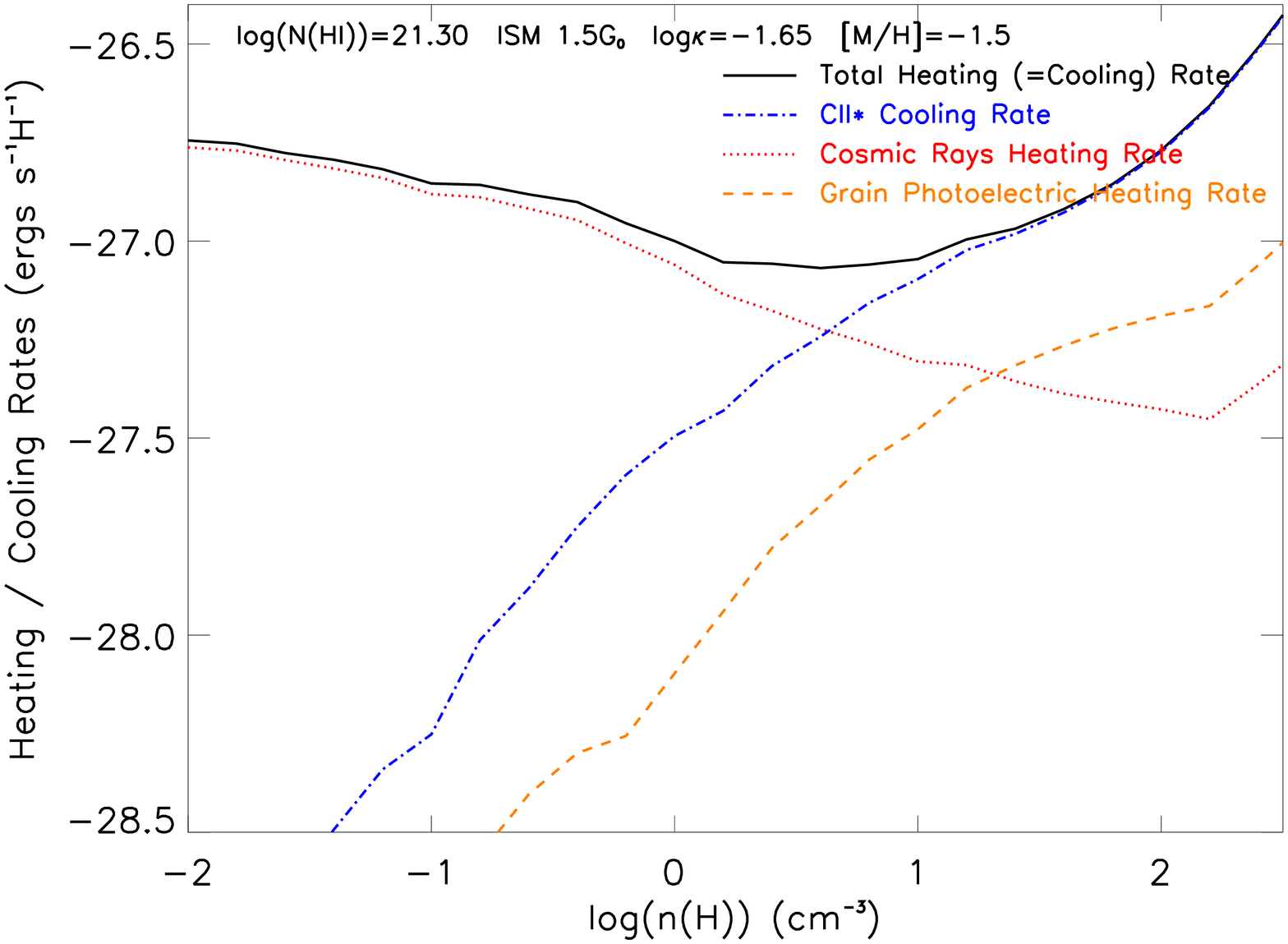}
\includegraphics[width=0.3\textwidth, angle=90]{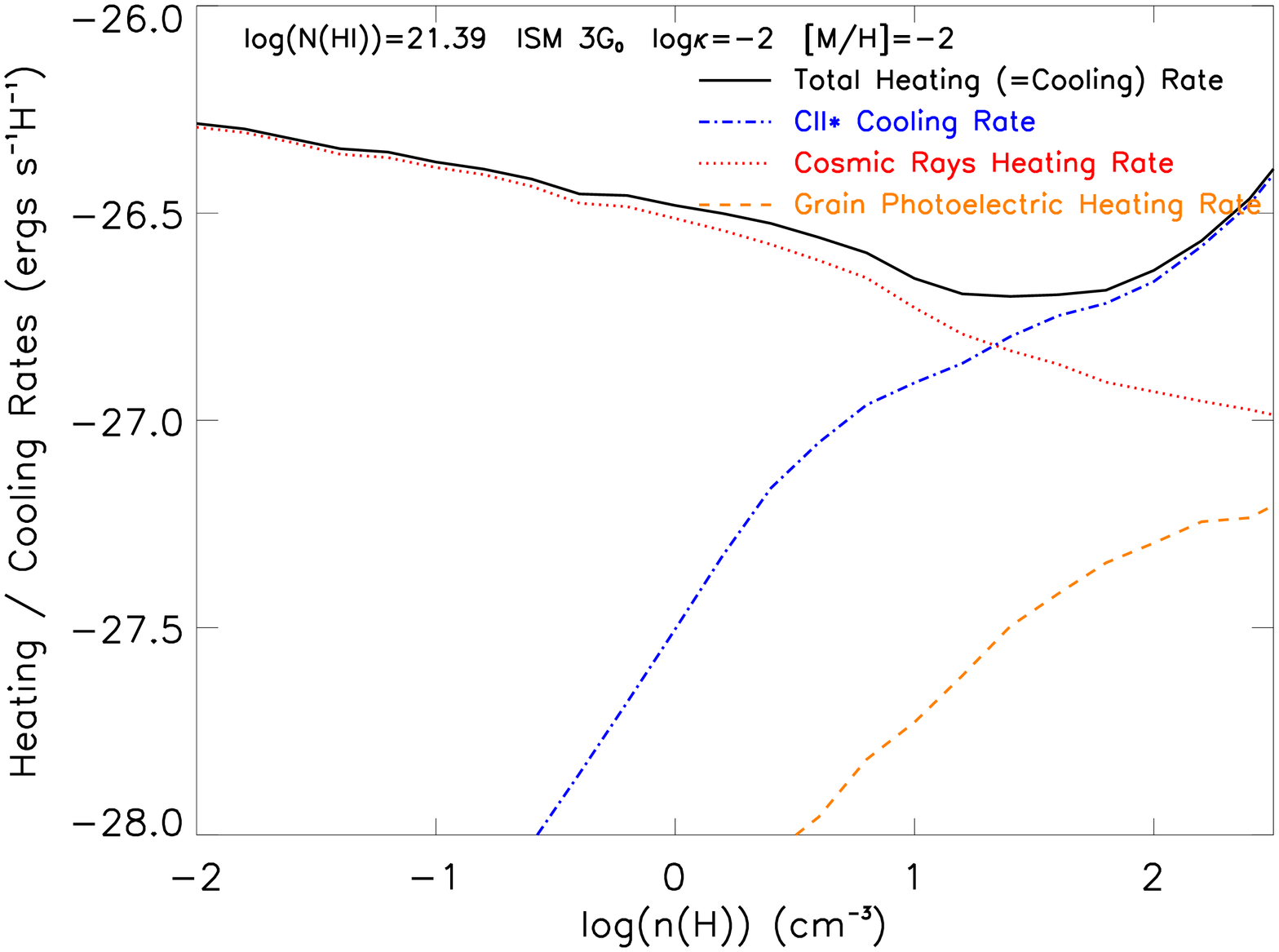}
\caption{Heating and cooling rates from our photoionization models for the DLAs at z$_{abs}$ = 2.5397 (top) and 2.68537 (bottom) towards J1004$+$0018.}
\label{fig:rates}
\end{figure}

We construct the phase diagrams for the DLAs (gas pressure vs. hydrogen number density n(H); see panels (a) of 
Figs.~\ref{fig:dla1} and \ref{fig:dla2}), by varying the n(H) and stopping the simulations when \textit{N}($\hon$) 
becomes equal to that observed in the DLAs. Results for  Haardt \& Madau radiation field as well as the local radiation 
field scaled by different scale factors are also plotted.
We put constraints on n(H) of the DLAs using the observed $\cto$* cooling rate and \textit{N}($\cto$*)/\textit{N}($\sto$) 
(panels (b) and (c) of Figs.\ref{fig:dla1} and \ref{fig:dla2} respectively). We use the 3-$\sigma$ upper 
limit of \textit{N}($\con$) as consistency checks on our models (panels (d) of Figs.~\ref{fig:dla1} and \ref{fig:dla2}).
From the results of the simulations, we observe that: 
\begin{enumerate}
\item For both systems, the extragalactic UV background radiation alone is not sufficient to produce the 
observed amount of $\cto$*, and a local radiation field (due to active star formation) is required to 
produce the  $\cto$* observations \citep[consistent with that found by][]{wolf03a,wolf08}.
\item The observed \textit{l$_{c}$} and \textit{N}($\cto$*)/\textit{N}($\sto$) are reproduced
with the ISM radiation field scaled by factor between 1 to 2 
and log~n(H) (cm$^{-3}$) = 0.7 $\pm$ 0.3 for DLA1, and the ISM field scaled by factor between 2 to 4
and log~n(H) (cm$^{-3}$) = 1.50 $\pm$ 0.20 for DLA2. As can be seen from the panels (a) of 
Figs.~\ref{fig:dla1} \& \ref{fig:dla2}, such conditions satisfy a stable CNM phase. 
Note that in typical VMP DLAs, C is slightly underabundant with respect to Si 
([C/Si] $\sim$ $-$0.2; see Fig. \ref{fig:comp}), while CLOUDY assumes solar composition. 
To account for this, we have included 0.2 dex in the upper bound on the observed 
\textit{N}($\cto$*)/\textit{N}($\sto$) for the DLA2 which is a low-metallicity DLA. 
\item We note that the above solutions are consistent with the observed upper limits of \textit{N}($\con$)/\textit{N}($\cto$*)
(see panels (d) Figs.~\ref{fig:dla1} \& \ref{fig:dla2}). The constraints could be better if we had high 
enough resolution to detect $\con$. Further we checked that our models do not produce any $\cfo$ and $\sif$.
While we do detect these high ions in our systems, they are shifted in velocity space with respect to the 
first ions and neutral species, indicating that they are most probably associated with a distinct phase \citep{fox07}. 
Our models also do not produce any $\sith$ and $\alth$. $\sith$ line is heavily blended in both the DLAs, 
and hence not an useful constraint. Our spectrum does not cover $\alth$ line for both the DLAs. Note that 
the presence of $\alth$ along with the absence of $\con$ would be inconsistent with $\cto$* absorption 
originating from the CNM \citep{sri05}. Hence detection of $\alth$ in these systems would imply that the 
$\cto$* absorption is arising from the warm/partially ionized gas. 
\item For both the above stable CNM solutions, heating due to cosmic rays dominates the total heating rate. 
This is demonstrated in Fig.~\ref{fig:rates}. In the top panel we show heating \& cooling rates from
the best-fitting photoionization model for the DLA1. It is clear from this figure that, in the density range 
constrained by the observations, $\cto$* dominates the net cooling rate and cosmic ray heating accounts for 
$\sim$ 60\% of the total heating rate. This is the case also for DLA2 (see bottom panel in Fig.~\ref{fig:rates}). 
The additional heating one sees for log~n(H) $\ge$ 2 is due to heating by collisional deexcitations 
of vibrationally excited H$_2$ molecules that are present in our models. This process can not be dominant in our 
systems as H$_2$ molecules are not detected. In summary, the observed $\cto$* in two DLAs in our sample are 
consistent with the gas being CNM, heated mainly by cosmic rays.
\end{enumerate}

\indent Higher radiation field is required to produce the observed $\cto$* absorption in DLA2 (having lower metallicity and 
dust-to-gas ratio) than in DLA1. However, converting this UV field into a star formation rate (SFR) is not straight
forward as UV field not only depends on SFR but also on the mean ISM opacity. In the low-metallicity, low-dust 
case we expect the mean free path for dust scattering to be larger and hence a higher UV radiation field even when
the SFR is low. However, we find that the main heating source in these DLAs are cosmic rays. We require
the CR background $1-2$ times and $2-4$ times that of what is seen in the the Milky Way for DLA1 and DLA2 respectively. 

The cosmic ray heating rate can be taken as proportional to the supernova rate, since cosmic rays are accelerated in 
supernova shock fronts, and hence related to the SFR \citep[see Eq. 9 of][]{wolf03a}. If we follow this argument,
then we can conclude that the present SFR in DLA1 and DLA2 are $1-2$ times and $2-4$ times that of the 
Milky Way respectively. Therefore, despite the gas phase metallicity being low in these systems, the required heating 
rate suggests that there is ongoing star formation activity in these systems. 

 Note that the CLOUDY models presented above assume
 one-sided illumination. We also ran test models assuming illumination
 from both the sides (using ``double'' command of CLOUDY) to check for
 differences. We find that that \textit{N}($\cto$*)/\textit{N}($\sto$) 
 and \textit{l$_{c}$} are slightly higher in the latter case, but the 
 differences are $\lesssim 0.1$ dex and within the errors of the observed 
 values of \textit{N}($\cto$*)/\textit{N}($\sto$) and \textit{l$_{c}$} 
 used to constrain the models. Hence the n(H) values and radiation field 
 that we obtained from our best-fitting models hold true in the latter case 
 also. 

 All our models assume constant density throughout the absorbing 
 cloud. As a check, we also ran constant pressure models for a few cases. 
 We find that the hydrogen density and temperature profiles do not change much 
 ($\lesssim 0.1$ dex) throughout the structure of the cloud, for initial 
 n(H) corresponding to the WNM and the CNM. However, for initial n(H) 
 corresponding to the unstable phase, the density and temperature drop 
 by $\sim1$ dex across the cloud. As shown earlier, the observed quantities 
 in the DLAs are well reproduced by stable CNM models. Then for initial
 n(H) corresponding to the average n(H) of the CNM models, there are no 
 significant changes in the results of the constant pressure models from 
 that of the constant density models. 

 Note that in all our model predictions, the column densities are obtained
 assuming our line of sight is perpendicular to the plane parallel slab.
 To check for inclination effects, we ran models assuming that the absorbing 
 slab is tilted with respect to the sightline (i.e., \textit{N}($\hon)^{\perp}$ 
 = $cos(\theta)~\times$ \textit{N}($\hon$) observed, with $\theta$ being the inclination angle). 
 In order to consider this, we took the stopping \textit{N}($\hon$) of the slab 
 as the observed value multiplied by the average value of cosine (i.e., 0.5).
 We find that this reduction in \textit{N}($\hon$) has no effect on the results 
 of the models, and cosmic rays are still the dominant heating agents required 
 to explain the observed quantities in the DLAs.

\begin{table*} 
\caption{Measurements of $\cto$* and cooling rates in DLAs in our sample}
\centering
\begin{tabular}{ccccccccc}
\hline
QSO          & z$_{abs}$ & log~\textit{N}($\hon$) & log~\textit{N}($\cto$*) & log~\textit{l$_{c}$}     & [Fe/H]        & [M/Fe]       & M  \\
             &           & (cm$^{-2}$)            & (cm$^{-2}$)             & (ergs s$^{-1}$ H$^{-1}$) &               &              &    \\
\hline 
J0035$-$0918 & 2.34006   & 20.55 (0.10)           & $\le$ 12.30             & $\le$ $-$27.77           & $-$2.98 (0.10) & 0.29 (0.09) & Si \\
J0234$-$0751 & 2.31815   & 20.90 (0.10)           & $\le$ 12.37             & $\le$ $-$28.05           & $-$2.22 (0.10) & 0.38 (0.06) & S  \\
J0953$-$0504 & 4.20287   & 20.55 (0.10)           & $\le$ 12.95             & $\le$ $-$27.12           & $-$2.98 (0.21) & 0.28 (0.19) & Si \\
J1004$+$0018 & 2.53970   & 21.30 (0.10)           & 13.63 (0.02)            & $-$27.19 (0.10)          & $-$1.67 (0.10) & 0.34 (0.02) & S  \\
J1004$+$0018 & 2.68537   & 21.39 (0.10)           & 13.94 (0.03)            & $-$26.97 (0.10)          & $-$2.18 (0.11) & 0.37 (0.03) & S  \\
J1004$+$0018 & 2.74575   & 19.84 (0.10)           & $\le$ 12.28             & $\le$ $-$27.08           & $-$2.03 (0.10) & 0.35 (0.02) & Si \\
\hline
\end{tabular}
\label{tab:ctos1}
\end{table*}
\begin{table*}
\caption{List of metal-poor DLAs in literature with $\cto$* detection}
\centering
\begin{tabular}{ccccccccc}
\hline
QSO                & z$_{abs}$ & log~\textit{N}($\hon$) & log~\textit{N}($\cto$*) & log~\textit{l$_{c}$}     & [Fe/H]         & [M/Fe]        & log~\textit{N}($\cto$*)/\textit{N}($\sto$) & Ref. \\
                   &           & (cm$^{-2}$)            & (cm$^{-2}$)             & (ergs s$^{-1}$ H$^{-1}$) &                &               &                                            &      \\
\hline
J001328.21$+$13582 & 3.2811    & 21.55 (0.15)           & 13.67 (0.04)            & $-$27.40 (0.16)          & $-$2.72 (0.02) & 0.62 (0.16)   & $-$0.90 (0.07)                             & 1    \\
SDSS0127$-$00      & 3.7274    & 21.15 (0.10)           & 13.20 (0.06)            & $-$27.47 (0.12)          & $-$2.90 (0.02) & 0.50 (0.10)   & $\le$ $-$0.72                              & 1    \\
PSS0133$+$0400     & 3.6919    & 20.70 (0.10)           & 12.95 (0.03)            & $-$27.27 (0.11)          & $-$2.74 (0.05) & 0.40 (0.16)   & $\le$ $-$0.92                              & 1    \\
J0255$+$00         & 3.9146    & 21.30 (0.05)           & 13.44 (0.04)            & $-$27.38 (0.06)          & $-$2.05 (0.09) & 0.27 (0.09)   & $-$1.28 (0.04)                             & 1    \\
J0311$-$1722       & 3.7340    & 20.30 (0.06)           & 13.55 (0.06)            & $-$26.27 (0.08)          & $\le$ $-$2.01  & $\ge$ $-$0.49 & 0.63 (0.09)                                & 2    \\
FJ0747$+$2739      & 3.9000    & 20.50 (0.10)           & 13.35 (0.07)            & $-$26.67 (0.12)          & $-$2.45 (0.03) & 0.44 (0.03)   & $-$0.29 (0.07)                             & 1    \\
PC0953$+$47        & 4.2442    & 20.90 (0.15)           & 13.60 (0.10)            & $-$26.82 (0.18)          & $-$2.52 (0.08) & 0.33 (0.09)   & $-$0.25 (0.10)                             & 1    \\
J120802.65$+$63032 & 2.4439    & 20.70 (0.15)           & 13.55 (0.03)            & $-$26.67 (0.15)          & $-$2.55 (0.01) & 0.23 (0.15)   & 0.05 (0.04)                                & 1    \\
Q1337$+$11         & 2.7959    & 20.95 (0.10)           & 13.11 (0.10)            & $-$27.36 (0.14)          & $-$2.03 (0.08) & 0.31 (0.14)   & $-$1.16 (0.11)                             & 1    \\
J1340$+$1106       & 2.7958    & 21.00 (0.06)           & 12.99 (0.04)            & $-$27.53 (0.07)          & $-$2.15 (0.06) & 0.32 (0.09)   & $-$1.30 (0.04)                             & 2    \\
BRI1346$-$03       & 3.7358    & 20.72 (0.10)           & 12.55 (0.11)            & $-$27.69 (0.15)          & $-$2.63 (0.02) & 0.30 (0.02)   & $-$1.01 (0.11)                             & 1    \\
J203642.29$-$05530 & 2.2805    & 21.20 (0.15)           & 13.36 (0.08)            & $-$27.36 (0.17)          & $-$2.24 (0.02) & 0.53 (0.17)   & $-$1.25 (0.11)                             & 1    \\
J231543.56$+$14560 & 3.2729    & 20.30 (0.15)           & 13.55 (0.08)            & $-$26.27 (0.17)          & $-$2.03 (0.03) & 0.25 (0.15)   & $-$0.09 (0.09)                             & 1    \\
\hline
\end{tabular}
\label{tab:ctos2}
\begin{flushleft}
$^{1}$ \citet[Table 1]{wolf08}; $^{2}$ \citet[Table 11]{cookb11}
\end{flushleft}
\end{table*}
\subsection{$\cto$* in low-metallicity DLAs}
\indent In Table \ref{tab:ctos1}, we give the limits or in case of detection the values of column densities and cooling rates of $\cto$* 
measured for the DLAs in our sample. We found a total 39 DLAs in literature with [Fe/H] $\le$ $-$2.0 for which the column density of $\cto$* 
or an upper limit to it has been measured \citep{wolf03a,sri05,wolf08,not08,cookb11}. In 26 systems $\cto$* is not detected and we summarize 
the basic properties of the remaining 13 systems in Table~\ref{tab:ctos2}. We notice that there is no difference in the distribution of 
metallicity between the systems with and without $\cto$*. However, systems with $\cto$* detections tend to have higher $N$(H~{\sc i}) 
(with a median value of log[$N$(H~{\sc i})(cm$^{-2}$)] = 20.9), compared to those without $\cto$* detection (with a median value of 
log[$N$(H~{\sc i})(cm$^{-2}$)] = 20.5).

From Table~\ref{tab:ctos2}, we note that 8 of the 13 DLAs from the literature 
sample, and all the systems in our sample (within errors) fall in the 
``low cool'' category of the bimodal distribution as defined by \citet{wolf08}.
The fraction of ``low cool'' DLAs we find among the $\cto$* detections is
much higher than that seen in the DLA sample of \citet{wolf08}. It is also
clear from the table that the typical value of log $N$(C~{\sc ii}*)/$N$(S~{\sc ii})
is $-$1.0 and $\ge$ $-$0.3 respectively for the ``low cool'' and ``high cool''
DLAs respectively.

\begin{figure}
\centering
\includegraphics[totalheight=0.3\textheight, angle=90]{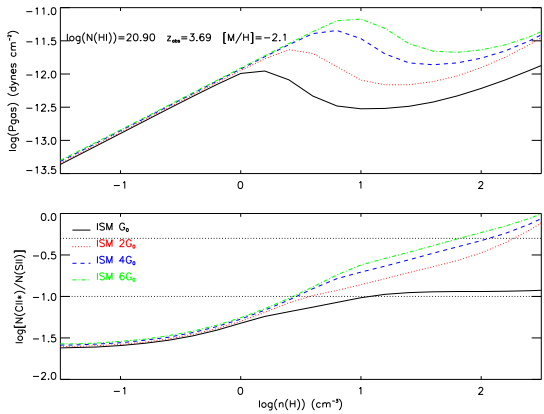}
\caption{Results of photoionization model having the median property of
the low-metallicity DLAs listed in Table~\ref{tab:ctos2}. Top panel 
shows the phase diagram and in the bottom panel 
log~$N$(C~{\sc ii}*)/$N$(S~{\sc ii}) is plotted against hydrogen density.}
\label{fig:med}
\end{figure}

To draw some broad conclusions we construct photo-ionization models 
considering the median properties of the DLAs in Table~\ref{tab:ctos2}.
We use log~$N$(H~{\sc i}) = 20.90, [M/H] = $-$2.2 and log~$\kappa$ = $-$2.4
in CLOUDY. The results for four values of the ionizing radiation are
given in Fig.~\ref{fig:med}. In the bottom panel we show the model
predicted values of log~$N$(C~{\sc ii}*)/$N$(S~{\sc ii}). It is clear
from the figure that log~$N$(C~{\sc ii}*)/$N$(S~{\sc ii}) $\sim$ $-$1 is
obtained for radiation field (and CR ionization rate) similar to 
what is seen in the Galaxy with n(H~{\sc i}) $\sim$ 10 cm$^{-3}$.
For log~$N$(C~{\sc ii}*)/$N$(S~{\sc ii}) $\ge$ $-$0.3 as found in the
case of ``high cool'' metal poor DLAs, one requires the radiation field
(and CR ionization rate) to be at least more than 4 times 
that seen in the Galaxy and n(H~{\sc i}) $\ge$ 60 cm$^{-3}$.
We also find in all these cases that the CR-heating dominates the 
gas heating rate. If we use a simple relation between SFR and 
CR ionization rate then the star formation rates in these metal poor
DLAs are similar to or higher than what is seen in the Milky Way.
In the models discussed above we use [C/S] as in solar composition.
From Table~11 of \citet{cookb11}, we can see [C/Si] is $\sim$ $-$0.20 in 
DLAs with carbon detection. If indeed C is slightly underabundant
compared to S then we will require higher density and higher
CR ionization rate compared to what we derive above. 
Based on the observed high J excitations of H$_{2}$ in DLAs, it has been 
inferred that UV radiation field similar to or higher than the mean Galactic 
UV background is required in these systems \citep{hira05,sri05}. However, these 
systems usually have at least an order of magnitude higher metallicity and dust 
compared to the low-metallicity DLAs \citep[see for e.g.][]{petj06}. Hence, 
irrespective of the metallicity, it seems that the inferred background radiation is 
similar to or higher than that of the Galaxy, whenever one uses indirect tracers 
to study the physical conditions in DLAs. However, the most challenging task is to 
convert this background radiation to the star formation rate. 

The requirement of ongoing star formation at a reasonable rate in a 
very metal poor DLA is similar to that required to explain the observations 
of the blue compact dwarf galaxy I Zw 18 in the local universe
\citep{Annibali13}. In this case it has been suggested that either
the star formation  is less efficient in populating the 
galaxy with metals or galactic winds have removed metals 
created in previous cycles of star formation. The same can be
applicable to the high-z low metal DLAs showing signatures of
ongoing star formation. Detecting the host galaxies
of DLAs will be interesting to study the nature of star formation
and magnetic fields (required for the generation and confinement
of CRs) in very metal poor gas at high-z.
%
%
\section{Summary \& Conclusions}  
\label{con} 
We have studied a sample of five DLAs and one sub-DLA and obtained their elemental abundances. Four of these DLAs and 
the sub-DLA turn out to be metal-poor, with [Fe/H] $\lesssim$ $-$2.0, and among these DLAs, two are extremely metal-poor, 
with [Fe/H] $\simeq$ $-$3.0. One of the extremely metal-poor systems (z$_{abs}$ = 2.34006 towards J0035$-$0918) studied here 
has also been analysed by \citet{cooka11}. This is also one of the two instances of CEMP DLAs that have been detected. 
However, we find that the extremely large enhancement of C in this system, as reported by \citet{cooka11}, is most probably 
due to the $b$ parameter being constrained to be the same for both the high and low mass ions, or in other words neglecting 
the thermal broadening. From our curve of growth analysis and best fit to the metal line profiles using VPFIT, we come 
to the conclusion that the line widths are likely to be determined by thermal broadening. In this case, carbon is still enhanced 
over iron, but by about three times. We estimate the DLA to have temperature in the range of 5000 $-$ 8000 K. Using this and 
from our CLOUDY model, we find that the physical conditions in this gas are consistent with a WNM in ionization equilibrium
with the metagalactic UV radiation field dominated by the QSOs and galaxies. \\ 
\indent The system at z$_{abs}$ = 4.20287 towards J0953$-$0504 is the most metal-poor DLA at z $>$ 4 studied till date. It shows
no enhancement of carbon or oxygen and its relative elemental abundance pattern is similar to that of a typical metal-poor DLA,
which may be explained by standard Population II nucleosynthesis results. The O/Fe, N/O and C/O ratios measured for our full 
sample conform to the observed trends seen in other metal-poor DLAs and stars. \\
\indent In addition, we have studied the cooling and heating processes in two of the DLAs showing $\cto$* absorption using CLOUDY.
The $\cto$* absorption is found to arise in the CNM phase and accounts for $\gtrsim$ 80\% of the total cooling in the DLAs. 
For the CNM solution, heating by cosmic rays is found to contribute $\sim$ 60\% to the total heating in the DLAs. Further, we 
find that an intergalactic background radiation field is not sufficient to produce the observed $\cto$* cooling in the DLAs, 
implying in-situ star formation in the DLAs, which is the same conclusion as reached by \citet{wolf06} and \citet{wolf08}. 
We infer the radiation field (in particular the CR ionization rate) in the DLAs to be similar to the local interstellar medium 
with amplitude scaled by some factor (G$_{0}$ ranging from 1 to 4 times that of the Galaxy). \\
\indent Lastly, we have assembled a sample of DLAs with [Fe/H] $\le$ $-$2.0 and showing $\cto$* absorption from the literature. 
Using photo-ionization models we argue that in such low-metal and low-dust systems, heating by grains is not as effective as in the 
Galactic ISM, and cosmic rays are most probably responsible for the observed excitation of $\cto$*. The inferred CR ionization rate 
varies from same as (in case of ``low cool'' metal poor DLAs) to more than four times (in case of ``high cool'' metal poor DLAs) that 
seen in the Milky Way. Understanding this high CR-ionizing rate in the metal poor DLAs will enable us to understand star formation
history, role of infall/winds and magnetic fields in low-metallicity galaxies in the early universe.
%
%

\noindent 
\section{ACKNOWLEDGEMENTS} 
We thank the editor and the anonymous referee for their useful comments.
Results presented in this work are based on observations carried out at the ESO under Prgm. ID. 383.A-0272 and 
086.A-0204 (PI: P. Petitjean) with the UVES spectrograph installed on the VLT at Cerro Paranal, Chile. This research has 
made use of the Keck Observatory Archive (KOA), which is operated by the W. M. Keck Observatory and the NASA Exoplanet 
Science Institute (NExScI), under contract with the National Aeronautics and Space Administration. RS and PPJ gratefully 
acknowledge support from the Indo-French Centre for the Promotion of Advanced Research (Centre Franco-Indien pour la 
Promotion de la Recherche Avanc\'{e}e) under contract No. 4304-2
%
%
\def\aj{AJ}%
\def\actaa{Acta Astron.}%
\def\araa{ARA\&A}%
\def\apj{ApJ}%
\def\apjl{ApJ}%
\def\apjs{ApJS}%
\def\ao{Appl.~Opt.}%
\def\apss{Ap\&SS}%
\def\aap{A\&A}%
\def\aapr{A\&A~Rev.}%
\def\aaps{A\&AS}%
\def\azh{AZh}%
\def\baas{BAAS}%
\def\bac{Bull. astr. Inst. Czechosl.}%
\def\caa{Chinese Astron. Astrophys.}%
\def\cjaa{Chinese J. Astron. Astrophys.}%
\def\icarus{Icarus}%
\def\jcap{J. Cosmology Astropart. Phys.}%
\def\jrasc{JRASC}%
\def\mnras{MNRAS}%
\def\memras{MmRAS}%
\def\na{New A}%
\def\nar{New A Rev.}%
\def\pasa{PASA}%
\def\pra{Phys.~Rev.~A}%
\def\prb{Phys.~Rev.~B}%
\def\prc{Phys.~Rev.~C}%
\def\prd{Phys.~Rev.~D}%
\def\pre{Phys.~Rev.~E}%
\def\prl{Phys.~Rev.~Lett.}%
\def\pasp{PASP}%
\def\pasj{PASJ}%
\def\qjras{QJRAS}%
\def\rmxaa{Rev. Mexicana Astron. Astrofis.}%
\def\skytel{S\&T}%
\def\solphys{Sol.~Phys.}%
\def\sovast{Soviet~Ast.}%
\def\ssr{Space~Sci.~Rev.}%
\def\zap{ZAp}%
\def\nat{Nature}%
\def\iaucirc{IAU~Circ.}%
\def\aplett{Astrophys.~Lett.}%
\def\apspr{Astrophys.~Space~Phys.~Res.}%
\def\bain{Bull.~Astron.~Inst.~Netherlands}%
\def\fcp{Fund.~Cosmic~Phys.}%
\def\gca{Geochim.~Cosmochim.~Acta}%
\def\grl{Geophys.~Res.~Lett.}%
\def\jcp{J.~Chem.~Phys.}%
\def\jgr{J.~Geophys.~Res.}%
\def\jqsrt{J.~Quant.~Spec.~Radiat.~Transf.}%
\def\memsai{Mem.~Soc.~Astron.~Italiana}%
\def\nphysa{Nucl.~Phys.~A}%
\def\physrep{Phys.~Rep.}%
\def\physscr{Phys.~Scr}%
\def\planss{Planet.~Space~Sci.}%
\def\procspie{Proc.~SPIE}%
\let\astap=\aap
\let\apjlett=\apjl
\let\apjsupp=\apjs
\let\applopt=\ao
\bibliographystyle{mn}
\bibliography{mybib}
\end{document}